\documentclass[12pt]{article}
  
\evensidemargin=\oddsidemargin
\usepackage{graphicx}
\usepackage{amsmath, amssymb, amsthm, mathrsfs, mathtools, amscd}
\usepackage{color}
\usepackage{enumerate}
\usepackage{times}
\usepackage{cite}
\usepackage[normalem]{ulem}
\newtheorem{Theorem}{Theorem}[section]
\newtheorem{Proposition}[Theorem]{Proposition}
\newtheorem{Corollary}[Theorem]{Corollary}
\newcommand{\benum}{\begin{enumerate}[{\rm (R10)}]}
\newcommand{\eenum}{\end{enumerate}}
 \newcommand{\bTheorem}{\begin{Theorem}}
\newcommand{\eTheorem}{\end{Theorem}}
\newcommand{\bProposition}{\begin{Proposition}}
\newcommand{\eProposition}{\end{Proposition}}
\newcommand{\bCorollary}{\begin{Corollary}}
\newcommand{\eCorollary}{\end{Corollary}}
\newcommand{\bProof }{\begin{proof}}
\newcommand{\eProof}{\end{proof}}
\newcommand{\beq}{\begin{equation}}
\newcommand{\eeq}{\end{equation}}
\newcommand{\beql}[1]{\begin{equation}\label{eq:#1}}
\newcommand{\beqa}{\begin{eqnarray}}
\newcommand{\eeqa}{\end{eqnarray}}
\newcommand{\beqas}{\begin{eqnarray*}}
\newcommand{\eeqas}{\end{eqnarray*}}
\newcommand{\deqs}[1]{ \begin{align*}#1\end{align*}}
\newcommand{\deq}[1]{ \begin{align}#1\end{align}}
\newcommand{\av}[1]{\langle#1\rangle}
\newcommand{\bracket}[1]{\langle#1\rangle}
\newcommand{\mb}{\mbox}
\newcommand{\nn}{\nonumber}
\newcommand\eg{{e.g.}}
\newcommand{\ie}{{i.e.}}
\newcommand{\C}{\mathbb{C}}
\newcommand{\N}{{\mathbb N}}
\newcommand{\Q}{{\mathbb Q}}
\newcommand{\R}{{\mathbb R}}
\newcommand{\al}{\alpha}
\newcommand{\be}{\beta}
\newcommand{\de}{\delta}
\newcommand{\ep}{\epsilon}
\newcommand{\ga}{\gamma}
\newcommand{\la}{\lambda}
\newcommand{\ph}{\phi}
\newcommand{\ps}{\psi}
\newcommand{\si}{\sigma}
\newcommand{\ta}{\tau}
\newcommand{\vp}{\varphi}
\newcommand{\De}{\Delta}
\newcommand{\Ga}{\Gamma}
\newcommand{\Eq}[1]{Eq.~(\ref{eq:#1})}
\newcommand{\eq}[1]{(\ref{eq:#1})}
\newcommand{\ZFC}{{\rm ZFC}}
\newcommand{\dom}{{\rm dom}}
\newcommand{\rank}{\mbox{\rm rank}}
\renewcommand{\And}{\wedge}
\newcommand{\Not}{\neg}
\newcommand{\Or}{\vee}
\newcommand{\IFF}{\Leftrightarrow}
\newcommand{\Iff}{\leftrightarrow}
\newcommand{\Inf}{\bigwedge}
\newcommand{\Sup}{\bigvee}
\newcommand{\THEN}{\Rightarrow}
\newcommand{\Then}{\rightarrow}
\newcommand{\Implies}{\rightarrow}

\renewcommand{\inf}{\bigwedge}
\renewcommand{\sup}{\bigvee}
\renewcommand{\iff}{\quad\mbox{iff}\quad}
\newcommand{\VQ}{V^{(\cQ)}}
\newcommand{\VL}{V^{(\cR)}}
\newcommand{\VB}{V^{(\cB)}}
\newcommand{\RQ}{R^{(\cQ)}}
\newcommand{\PRQ}{\cP(\ck{\Q})^{(\cQ)}}
\newcommand{\PXQ}{\cP(\ck{X})^{(\cQ)}}
\newcommand{\SAH}{{\rm SA}({\mathcal H})}
\newcommand{\cLin}{{\cL(\in)}}
\newcommand{\val}[1]{[\![#1]\!]}
\newcommand{\valj}[1]{[\![#1]\!]_{j}}
\newcommand{\valR}[1]{\val{#1}_{\cR}}
\newcommand{\valQ}[1]{\val{#1}_{\cQ}}
\newcommand{\cdom}{\rotatebox[origin=c]{90}{$\models$}}
\newcommand{\commutes}{\,\rotatebox[origin=c]{270}{$\multimap$}\,}
\newcommand{\cuniv}{\underline{\Or}}
\newcommand{\cm}{\underline{\Or}}
\newcommand{\com}{\rotatebox[origin=c]{90}{$\models$}}
\newcommand{\p}{{}^{\perp}}
\newcommand{\ck}[1]{\check{#1}}
\newcommand{\id}{{\rm id}}
\newcommand{\les}{\preccurlyeq}
\newcommand{\cA}{{\mathcal A}}
\newcommand{\cB}{{\mathcal B}}
\newcommand{\cC}{{\mathcal C}}
\newcommand{\cD}{{\rm dom}}
\newcommand{\cF}{{\mathcal F}}
\newcommand{\cH}{{\mathcal H}}
\newcommand{\cI}{{\mathcal I}}
\newcommand{\cL}{{\mathcal L}}
\newcommand{\cM}{{\mathcal M}}
\newcommand{\cP}{{\mathcal P}}
\newcommand{\cQ}{{\mathcal Q}}
\newcommand{\cR}{{\mathcal R}}
\newcommand{\cS}{{\mathcal S}}
\newcommand{\hA}{\hat{A}}
\newcommand{\hB}{\hat{B}}
\newcommand{\tP}{\tilde{P}}
\newcommand{\tQ}{\tilde{Q}}
\begin{document}
\title{\bf Quantum Set Theory: \\ Transfer Principle and De Morgan's Laws\thanks{
 A preliminary account of this research is to appear
in the Proceedings of the Symposium on Advances in Mathematical Logic 2018
(SAML 2018), Kobe, Japan, September 18--20.}
 \\[10pt]
{\normalsize\it Dedicated to the memory of Professor Gaisi Takeuti}
 }
 \author{\sc Masanao Ozawa\\ \\
{\it\small
College of Engineering,
Chubu University,
1200 Matsumoto-cho, Kasugai 487-8501, Japan}\\
{\small\it Graduate School of Informatics,
Nagoya University,
Chikusa-ku, Nagoya 464-8601, Japan}}
\date{}

\maketitle
\begin{abstract}
In quantum logic, introduced by Birkhoff and von Neumann, De Morgan's Laws play an important role in the projection-valued truth value assignment of observational propositions in quantum mechanics. Takeuti's quantum set theory extends this assignment to all the set-theoretical statements on the universe of quantum sets. However, Takeuti's quantum set theory has a problem in that De Morgan's Laws do not hold between universal and existential bounded quantifiers.  Here, we solve this problem by introducing a new truth value assignment for bounded quantifiers that satisfies De Morgan's Laws. To justify the new assignment, we prove the Transfer Principle, showing that this assignment of a truth value to every bounded ZFC theorem has a lower bound determined by the commutator, a projection-valued degree of commutativity, of constants in the formula. We study the most general class of truth value assignments and obtain necessary and sufficient conditions for them to satisfy the Transfer Principle, to satisfy De Morgan's Laws, and to satisfy both.  For the class of assignments with polynomially definable logical operations, we determine exactly 36 assignments that satisfy the Transfer Principle and exactly 6 assignments that satisfy both the Transfer Principle and De Morgan's Laws.
\medskip

\newcommand{\sep}{,\ }
\medskip

 \end{abstract}
\newpage

\tableofcontents
\newpage

\section{Introduction}\label{se:1}
Quantum set theory originated from the methods of forcing introduced 
by Cohen \cite{Coh63,Coh66} for independence proofs and from the quantum logic
introduced by Birkhoff and von Neumann \cite{BvN36}.
It relates two remote fields of mathematics: 
the foundations of mathematics, and the foundations of quantum mechanics.
After Cohen's work, Scott and Solovay \cite{SS67} reformulated 
the method of forcing in terms of Boolean-valued models of set theory \cite{Bel05},
which incorporates various extensions of the notion of sets,
such as sheaves \cite{FS79}, topoi \cite{Joh77}, 
and intuitionistic set theory \cite{Gra79}. 
As a successor of the above attempts, 
Takeuti \cite{Ta81} introduced quantum set theory, a set theory based on the 
quantum logic of Birkhoff and von Neumann.

Birkhoff and von Neumann \cite{BvN36} introduced quantum logic as the logic of 
observational propositions in quantum mechanics and revealed that quantum indeterminacy
is closely related to the violation of the distributive law in quantum logic,
whereas the logic of classical mechanics obeys the laws of classical logic including the distributive law.
In his seminal paper \cite{Ta81},
Taketui proposed the program of developing mathematics based on quantum logic,
in a way similar with developing constructive mathematics based on intuitionistic logic,
specifically, by constructing set theory based on quantum logic.  Takeuti pointed
out the difficulty of the program due to the drastic difference of quantum logic
from classical or intuitionistic logic, whereas he envisaged the richness of the mathematics
based on quantum logic, since the semantical structure of quantum logic is made up
of pasted many Boolean subalgebras, analogously with the fact that the space-time 
geometry of the relativity theory is locally homeomorphic to the classical, flat geometry. 

Takeuti also showed that the real numbers internally constructed in quantum set 
theory are externally in one-to-one correspondence with the physical quantities, called the observables, 
of the quantum system associated with that logic, or, equivalently, with the self-adjoint 
operators on the associated Hilbert space.  According to von Neumann's axiomatization 
of quantum mechanics \cite{vN32E}, the laws of quantum mechanics are derived from the
structure of self-adjoint operators on a Hilbert space, and hence the above result
demonstrated a remarkable aspect of Takeuti's program that quantum mechanics can be
viewed as the real number theory internally developed in quantum set theory.  
From this point of view, quantum set theory is expected to make significant contributions
to the foundations of quantum mechanics and applications of quantum mechanics
including quantum information and quantum computing.

\sloppy
To be more precise, 
Takeuti constructed the universe $\VQ$ of quantum sets based on the 
quantum logic $\cQ$ represented by the lattice of projections on a Hilbert space $\cH$.
To every formula $\ph(x_1,\ldots,x_n)$ in set theory and to any quantum sets 
$u_1,\ldots,u_n$ in the universe $\VQ$, he assigned the $\cQ$-valued truth value 
$\val{\ph(u_1,\ldots,u_n)}$ of the formula $\ph(x_1,\ldots,x_n)$ to be
satisfied by  $u_1,\ldots,u_n\in\VQ$.
Due to the well-known arbitrariness of implication in quantum logic,
he adopted the Sasaki arrow for implication.
In order to provide quantum counterparts of the axioms of ZFC, he introduced 
the notion of the commutator of elements of the universe $\VQ$, a measure of the 
degree of commutativity, and he showed that the axioms of
ZFC hold in the universe $\VQ$ if appropriately modified by the commutators.
Based on his earlier work on Boolean-valued analysis \cite{Ta78},
he derived that the real numbers in the universe $\VQ$ 
correspond to the self-adjoint operators on the underlying Hilbert space $\cH$,
suggesting rich applications to quantum physics and analysis 

Following Takeuti's work, we explored the question of how the theorems of 
ZFC hold in the universe $\VQ$.
We showed that the following Transfer Principle holds for Takeuti's quantum
set theory \cite{07TPQ}.
\smallskip

{\bf Transfer Principle.}
{\em Every $\De_0$-formula $\ph(x_1,\ldots,x_n)$ in the language of set theory 
provable in ZFC holds for any elements $u_1,\ldots,u_n$ in the universe $\VQ$ 
with the $\cQ$-valued truth value $\val{\phi(u_1,\ldots,u_n)}$ 
at least the commutator $\cm(u_1,\ldots,u_n)$
of $u_1,\ldots,u_n\in\VQ$, \ie,
\[
\val{\phi(u_1,\ldots,u_n)}\ge \cm(u_1,\ldots,u_n).
\]
}
\indent
This result was extended to general complete orthomodular
lattices and to a general class of operations for implication \cite{17A2}.
Note that this generalization of the formulation unifies quantum set
theory with Boolean-valued models of set theory, which are included
as the case where $\cQ$ is a Boolean algebra, 
and naturally incorporates the methods of Boolean-valued analysis 
\cite{Sco69a,Ta78,
Ta79c,Ta83a,
Ta88,
Eda83,Jec85,
KK99,Nis91,83BT,
83BH,84CT,
85NC,
85TN,
94FN,95SB,
Smi84} 
into various applications of quantum set theory.
Quantum set theory has been applied to quantum mechanics to extend 
the probabilistic predictions from observational propositions to 
relations between observables, such as commutativity, equality, and order relations 
\cite{11QRM,16A2,17A1}, as well as being applied to computer science \cite{Yin05}.
Relations with paraconsistent set theory, intuitionistic set theory, 
and topos quantum mechanics have also been studied recently 
\cite{Eva15,DEO20,Tit18}. 

In spite of this development of the theory, 
one problem has eluded a solution: Takeuti's assignment 
of the truth value does not satisfy De Morgan's Laws for the 
universal--existential pair of bounded quantifiers.
Since the inception of quantum logic, due to 
Birkhoff and von Neumann \cite{BvN36},
interpretations of connectives have often been polemical,
but De Morgan's Laws have played an important role.
For instance, in quantum logic, the meaning of disjunction is less obvious
than that of conjunction and negation, 
and yet De Morgan's Laws enable us to determine disjunction 
in terms of conjunction and negation.

In the present paper, we examine Takeuti's truth value assignment 
of a  truth value $\val{\ph}$ in the quantum logic $\cQ$
to a set theoretic statement $\ph$.
In particular,  Takeuti noted
\begin{quote}
In Boolean-valued universes, 
$\val{(\forall x\in u)\ph(x)}=\val{\forall x(x\in u\Then \ph(x)}$ and 
$\val{(\exists x\in u)\ph(x)}=\val{\exists x(x\in u \And \ph(x)}$ [hold]. 
But this is not the case for $\VQ$. \cite[p.~315]{Ta81}
\end{quote} 
\noindent
and defined the truth values of bounded 
quantifications using the Sasaki arrow $\Then$ defined by 
$P\Then Q=P\p\Or(P\And Q)$
 as follows. 
\benum
\item[(1)] $ \val{(\forall x\in u)\, {\ph}(x)} 
= \Inf_{u'\in \dom(u)}
(u(u') \Then \val{\ph(u')})$.
\item[(2)] $ \val{(\exists x\in u)\, {\ph}(x)} 
= \Sup_{u'\in \dom(u)}
(u(u') \And \val{\ph(u')})$.
\eenum

However, there is a problem in that the classical implication 
$P\Then Q=P\p\Or Q$ was avoided in the bounded universal 
quantification, and yet the classical conjunction $\And$ was
used in the bounded existential quantification. 
Since 
the relation $P\And Q=(P\Then Q^{\perp})^{\perp}$ 
does not hold for the classical conjunction
$\And$ and the Sasaki arrow $\Then$, 
De Morgan's Laws,
\benum
\item[(3)] $ \val{\neg(\forall x\in u)\, {\ph}(x)}=
 \val{(\exists x\in u)\,\neg {\ph}(x)}$,
\item[(4)] $ \val{\Not(\exists x\in u)\, {\ph}(x)}
=  \val{(\forall x\in u)\, \neg{\ph}(x)}$,
\eenum
\noindent 
do not hold.  In fact, we shall show that there is a predicate $\ph(x)$ 
in Takeuti's quantum set theory such that $\val{(\exists x\in u)\,\neg {\ph}(x)}=0$ 
but $\val{\neg(\forall x\in u)\, {\ph}(x)}>0$.

In the present paper, we introduce a new binary operation $*$ by
$P*Q=(P\Then Q^{\perp})^{\perp}$ in quantum set theory
and redefine the truth values of 
membership relation and
bounded existential quantification as follows.
\benum
\item[(5)]  $ \val{u \in v} 
= \sup_{v' \in \cD(v)} (v(v') * \val{v'=u})$.
\item[(6)] $ \val{(\exists x\in u)\, {\ph}(x)} 
= \Sup_{u'\in \dom(u)}
(u(u') * \val{\ph(u')})$.
\eenum
\noindent
Then, De Morgan's Laws hold for bounded universal
quantification and bounded existential quantification.
Thus,  for the language of quantum set theory, we can assume 
only negation,
conjunction, and bounded and unbounded universal quantification as
primitive, while disjunction, bounded and unbounded existential 
quantification are considered to be introduced by definition.

The operation $\ast$ was found by Sasaki \cite{Sas54},
and has been studied  as the Sasaki projection in connection 
with residuation theory \cite{Har81}, 
whereas, as far	as our knowledge extends, this operation has not been used for 
defining bounded quantifiers in quantum logic.  

Because of the well-known arbitrariness of choosing the connective
for implication in quantum logic \cite{Urq83}, 
we previously introduced a general class of binary operations $\Then$ 
for implication
on a general quantum logic represented by a complete orthomodular lattice
\cite{17A2}.
In the present paper, we continue to explore these operations for the problem
of the consistency between the Transfer Principle and De Morgan's Laws.
For this purpose, we introduce another general class of binary operations $\ast$
for conjunction. 
Then, we will ask questions as to which pairs $(\Then,\ast)$ support the Transfer
Principle and which pairs $(\Then,\ast)$ support both the Transfer Principle
and De Morgan's Laws, and we will answer these questions.  
For polynomially definable operations, we determine all 36 pairs $(\Then,\ast)$
 that admit the Transfer Principle, and we determine the 6 out of these 36 
that admit both the Transfer Principle and De Morgan's Laws,
including
the pair of the Sasaki arrow $\Then$
and the Sasaki projection $\ast$ 
and also the pair of the classical implication $\Then$ 
and the classical conjunction $\ast$, 
as previously mentioned in~\cite{17A2}.

This paper is organized as follows.
Section \ref{se:QL} discusses general properties of quantum logic represented by
a general complete orthomodular lattice (COML) $\cQ$.
Section \ref{se:3} discusses quantization of  operations in classical logic
including 96 polynomially definable operations found by Kotas \cite{Kot67}
and also polynomially indefinable operations, which were 
introduced by Takeuti \cite{Ta81}
and extensively studied in Ref.~\cite{17A2}.
To formulate a sound general theory, we introduce a class of binary operations,
called local binary operations, on a general COML $\cQ$, which share two
local properties with the polynomially definable ones. 
Section \ref{se:4} studies quantum set theory based on the universe $\VQ$ constructed
on an arbitrary COML $\cQ$ and $\cQ$-valued interpretations,
$\cQ$-valued truth value assignments, $\cI(\Then,\ast)$,
determined by arbitrary pairs $(\Then,\ast)$ of local binary operations on $\cQ$.
We characterize all the $\cQ$-valued interpretations $\cI(\Then,\ast)$ that admit 
the Transfer Principle as well as those that admit both the Transfer Principle and
De Morgan's Laws.  For polynomially definable operations $\Then$ and $\ast$,
this result determines 6 $\cQ$-valued interpretations $\cI(\Then,\ast)$ 
that satisfy both the Transfer Principle and De Morgan's Laws.
We also discuss applications of the above results to the notion
of spectral order in operator theory.
Section \ref{se:5} concludes the present paper.
We also discuss new interpretations of quantum logical connectives 
using the commutator based direct product decomposition developed in
Section \ref{se:3}.

\section{Quantum Logic}\label{se:QL}
\subsection{Complete orthomodular lattices}
A {\em complete orthomodular lattice}  is a complete
lattice $\cQ$ with an {\em orthocomplementation},
a unary operation $\perp$ on $\cQ$ satisfying
(i)  if $P \le Q$, then $Q^{\perp}\le P^{\perp}$,
(ii) $P^{\perp\perp}=P$,
(iii) $P\Or P^{\perp}=1$ and $P\And P^{\perp}=0$,
where $0=\Inf\cQ$ and $1=\Sup\cQ$,
that  satisfies the {\em orthomodular law}:
if $P\le Q$, then $P\Or(P^{\perp}\And Q)=Q$.
In this paper, any complete orthomodular lattice is called a {\em logic}.

A non-empty subset of a logic $\cQ$ is called a {\em sublattice} iff
it is closed under meet $\And $ and join $\Or$.
A sublattice is called a {\em subalgebra} iff it is further closed under 
orthocomplementation $\perp$.
A sublattice or a subalgebra $\cR$ of $\cQ$ is said to be {\em complete} iff 
for each subset $\cA$ of $\cR$, its infimum $\Inf \cA$ and supremum $\Sup \cA$ are in $\cQ$.
For any subset $\cA$ of $\cQ$, 
the subalgebra generated by $\cA$ is denoted by
$\Ga_0\cA$, and the complete subalgebra 
generated by $\cA$ is denoted by $\Ga\cA$.
We refer the reader to Kalmbach \cite{Kal83} for a standard reference on
orthomodular lattices.

We say that $P$ and $Q$ in a logic $\cQ$
{\em commute}, in  symbols
$P\commutes Q$, iff  $P=(P\And Q)\Or(P\And
Q^{\perp})$. All the relations $P\commutes Q$,
$Q\commutes P$,
$P^{\perp}\commutes Q$, $P\commutes Q^{\perp}$,
and $P^{\perp}\commutes Q^{\perp}$ are equivalent.
The distributive law does not hold in general, but
the following useful proposition  holds \cite[pp.~24--25]{Kal83}.

\bProposition\label{th:distributivity}
If $P,Q\commutes E$, then the sublattice generated by $P,Q,E$ is distributive.
\eProposition

When applying a distributive law under the assumption of 
Proposition \ref{th:distributivity}, we shall say that we are
{\em focusing} on $E$.
From Proposition \ref{th:distributivity}, a logic $\cQ$ is a Boolean
algebra if and only if $P\commutes Q$  for all $P,Q\in\cQ$.
In this case,  logic $\cQ$ is called {\em Boolean}.

The following proposition is useful for later discussions
\cite[Proposition 3.4]{Kal83}; an elementary proof is given
for the reader's convenience.

\bProposition\label{th:logic}
If $P_{\al}, E\in\cQ$ and 
$P_{\al}\commutes E$ for all $\al$, then
\beqas
&(\Sup_{\al}P_{\al})\commutes E,\quad \Inf_{\al}P_{\al}\commutes E,\quad
(\Sup_{\al}P_{\al})\And E=\Sup_{\al}(P_{\al}\And E),&\\
&(\Inf_{\al}P_{\al})\And  E =\Inf_{\al}(P_{\al}\And E).&
\eeqas
\eProposition
\bProof 
Suppose that $P_{\al},E\in\cQ$ and $P_{\al}\commutes E$ for every $\al$.
From
\beq
\Sup_{\al}(P_\al\And  E)\le E,\quad
\Sup_{\al}(P_\al\And  E^\perp)\le E^{\perp},
\eeq
we have
\beqa\label{eq:sup-com}
\Sup_{\al}(P_\al\And  E)\commutes E,\quad
\Sup_{\al}(P_\al\And  E^\perp)\commutes E.
\eeqa
By assumption, we have
$P_{\al}=(P_\al\And  E)\Or(P_\al\And  E^\perp)$ for every $\al$.
Since 
\beqas
\Sup_{\al}P_{\al}&=&
\Sup_{\al}[(P_\al\And  E)\Or(P_\al\And  E^\perp)]
=
\Sup_{\al}(P_\al\And  E)\Or\Sup_{\al}(P_\al\And  E^\perp),
\eeqas
we conclude, using \Eq{sup-com}, that $\Sup_{\al}P_{\al}\commutes E$.
Focusing on $E$, we then have, by \Eq{sup-com},
\beqas
(\Sup_{\al}P_{\al})\And E
&=&
[\Sup_{\al}(P_\al\And E)\Or\Sup_{\al}(P_\al\And  E^\perp)]\And E
=
\Sup_{\al}(P_\al\And  E).
\eeqas
Thus, we conclude
$(\Sup_{\al}P_{\al})\And E=
\Sup_{\al}( P_\al \And E)$.
The rest of the assertions follow similarly.
\eProof

For any subset $\cA\subseteq\cQ$,
we denote by $\cA^{!}$ the {\em commutant} 
of $\cA$ in $\cQ$ \cite[p.~23]{Kal83}, \ie, 
$$
\cA^{!}=
\{P\in\cQ\mid P\commutes Q \mbox{ for all }
Q\in\cA\}.
$$
Then $\cA^{!}$ is a complete subalgebra of 
$\cQ$  by Proposition \ref{th:logic} and satisfies $\cA^{!!!}=\cA^{!}$.
A {\em sublogic} of $\cQ$ is a subset $\cA$ of
$\cQ$ satisfying $\cA=\cA^{!!}$. 
Thus, any sublogic of $\cQ$ is a complete subalgebra of $\cQ$.
A sublogic $\cA$ is called {\em Boolean} iff $P\commutes
Q$  for all $P,Q\in\cA$.

For any subset $\cA\subseteq\cQ$, the smallest 
logic including $\cA$ is the logic
$\cA^{!!}$, called the  {\em logic generated by
$\cA$}.
We have $\cA
\subseteq \Ga \cA\subseteq
\cA^{!!}$.
Then it is easy to see that a subset 
 $\cA$ is a Boolean sublogic, or equivalently 
 a distributive sublogic, if and only if
$\cA=\cA^{!!}\subseteq\cA^{!}$.
If $\cA\subseteq\cA^{!}$, the subset
$\cA^{!!}$ is the smallest Boolean sublogic including
$\cA$.
A maximal Boolean sublogic $\cB$ of $\cQ$ is characterized by
$\cB^!=\cB$.
By Zorn's lemma, for every subset $\cA$ of $\cQ$ consisting of
mutually commuting elements, there is a maximal Boolean
sublogic of $\cQ$ including  $\cA$.

For any logic $\cQ$, the set $\cQ^{!}$ is called the {\em center} of $\cQ$
and denoted by $Z(\cQ)$.
Since $Z(\cQ)\subseteq \cQ=Z(\cQ)^{!}$, the center of $\cQ$ is a Boolean sublogic.
For any subset $\cA$ of $\cQ$, the center of the logic $\cA^{!!}$ generated by $\cA$
 is given by $Z(\cA^{!!})=\cA^{!}\cap \cA^{!!}$.

\subsection{Commutators}\label{sse:CIQL}
The commutator $\cdom(P,Q)$ 
of two elements $P$ and $Q$ of a logic $\cQ$ was introduced 
by Marsden \cite{Mar70} as
\beql{comM}
\cdom(P,Q)=(P\And Q)\Or(P\And Q\p)\Or(P\p\And Q)\Or(P\p\And Q\p).
\eeq
This notion was generalized to finite subsets of $\cQ$ by Bruns \& Kalmbach 
\cite{BK73} as
\beql{comBK}
\cdom(\cF)=\Sup_{\theta:\cF\to\{\id,\perp\}}\Inf_{P\in\cF}P^{\,\theta(P)}
\eeq
for any finite subset $\cF$ of $\cQ$, where
$\{\id,\perp\}$ stands for the set consisting of the identity operation 
$\id$ and the orthocomplementation~$\perp$.
Generalizing the notion of commutator to arbitrary subsets $\cA$ 
of $\cQ$, Takeuti \cite{Ta81} defined the {\em commutator} 
$\cdom(\cA)$ of $\cA$ by
\beql{comT}
\cdom(\cA)=\Sup \{E\in\cA^{!} \mid P\And E\commutes Q\And E
\mb{ for all }P,Q\in\cA\}
\eeq
for any subset $\cA$ of $\cQ$, which is consistent with \Eq{comBK}
if $\cA$ is a finite subset \cite[Proposition 4]{Ta81}.
With Takeuti's definition, it is not clear whether the commutator $\cdom(\cA)$ 
is determined inside the logic $\cA^{!!}$ generated by $\cA$ or not, 
unlike the definition of $\cdom(\cF)$ for finite subsets $\cF$. 
To resolve this problem, we have proved
\beql{comO}
\cdom(\cA)=\max \{E\in\cA^{!}\cap \cA^{!!} \mid P\And E\commutes Q\And E
\mb{ for all }P,Q\in\cA\}
\eeq
for any subset $\cA$ of $\cQ$ \cite[Theorem 2.2]{16A2}.
From the above, we conclude $\cdom(\cA)\in \cA^{!}\cap \cA^{!!}$.
Since every central element $E$ in a logic $\cR$ leads to the direct product decomposition
$\cR=[0,E]\times [0,E\p]$ \cite[Theorem 1.1]{Kal83}, the above result leads to 
the following theorem \cite[Theorem 2.4]{16A2}.

\bTheorem[Decomposition Theorem]\label{th:decomposition}
Let $\cA$ be a subset of a logic $\cQ$.
Then the sublogic $\cA^{!!}$ generated by $\cA$ is isomorphic to the direct product 
of the complete Boolean algebra
$[0,\cdom(\cA)]_{\cA^{!!}}$ and the complete orthomodular lattice
$[0,\cdom(\cA)^\perp]_{\cA^{!!}}$ without non-trivial Boolean factor.
\eTheorem
We refer the reader to Pulmannov\'{a} \cite{Pul85} and Chevalier \cite{Che89} 
for further results about commutators in orthomodular lattices. 

\subsection{Logics on Hilbert spaces}
Let $\cH$ be a Hilbert space.
For any subset $S\subseteq\cH$,
we denote by $S^{\perp}$ the orthogonal complement
of $S$.
Then $S^{\perp\perp}$ is the closed linear span of $S$.
Let $\cC(\cH)$ be the set of all closed linear subspaces in
$\cH$. 
With the set inclusion ordering, 
the set $\cC(\cH)$ is a complete
lattice. 
The operation $M\mapsto M^\perp$ 
is  an orthocomplementation
on the lattice $\cC(\cH)$, with which $\cC(\cH)$ is a logic.

Denote by $\cB(\cH)$ the algebra of bounded linear
operators on $\cH$ and $\cQ(\cH)$ the set of projections on $\cH$.
We define the {\em operator ordering} on $\cB(\cH)$ by
$A\le B$ iff $(\ps,A\ps)\le (\ps,B\ps)$ for
all $\ps\in\cH$. 
For any $A\in\cB(\cH)$, denote by $\cR (A)\in\cC(\cH)$
the closure of the range of $A$, i.e., 
$\cR(A)=(A\cH)^{\perp\perp}$.
For any $M\in\cC(\cH)$,
denote by $\cP (M)\in\cQ(\cH)$ the projection operator 
of $\cH$
onto $M$.
Then $\cR\cP (M)=M$ for all $M\in\cC(\cH)$
and $\cP\cR (P)=P$ for all $P\in\cQ(\cH)$,
and we have $P\le Q$ if and only if $\cR (P)\subseteq\cR (Q)$
for all $P,Q\in\cQ(\cH)$,
so that $\cQ(\cH)$ with the operator ordering is also a logic
isomorphic to $\cC(\cH)$.
Any sublogic of $\cQ(\cH)$ will be called a {\em logic on $\cH$}. 
For any $P,Q\in\cQ(\cH)$, we have 
$P\commutes Q$ iff  $PQ=QP$.

For any $\cA\subseteq\cB(\cH)$,
we denote by $\cA'$ the {\em commutant of 
$\cA$ in $\cB(\cH)$}, \ie, \[\cA'=\{X\in\cB(\cH)\mid
XA=AX\mb{ for all } A\in\cA\}.\]
A self-adjoint subalgebra $\cM$ of $\cB(\cH)$ is called a
{\em von Neumann algebra} on $\cH$ iff 
$\cM''=\cM$.
For any self-adjoint subset $\cA\subseteq\cB(\cH)$,
$\cA''$ is the von Neumann algebra generated by $\cA$.
We denote by $\cQ(\cM)$ the set of projections in
a von Neumann algebra $\cM$.
Then a subset $\cQ \subseteq\cQ(\cH)$ is a logic on $\cH$ if,
and only if, $\cQ=\cQ(\cM)$ for some von Neumann algebra
$\cM$ on $\cH$ \cite[Proposition 2.1]{07TPQ}.
In this case, we have $\cQ=\cQ^{!!}=\cQ(\cQ'')$.

\section{Quantization of Logical Operations}\label{se:3}
\subsection{Local operations}
Let $\cQ$ be a logic.
A binary operation $f:\cQ^2\to \cQ$ is said to be {\em local}
iff the following conditions are satisfied.
\benum
\item[{\rm (L1)}] $f(P,Q)\in\{P,Q\}^{!!}$ for all $P,Q\in\cQ$.
\item[{\rm (L2)}] $f(P,Q)\And E=f(P\And E,Q\And E)\And E$ if $P, Q\commutes E$ for all 
$P,Q,E\in\cQ$.
\end{enumerate}
 
 Note that by property (L1)  every sublogic of a logic $\cQ$ is invariant 
 under any local binary operation on $\cQ$.  
 The following theorem is useful for later discussions. 
 \bTheorem \label{th:LO}
Let $f$ be a local binary operation on a logic $\cQ$.
Let $P_{\al}, Q_{\al}, E\in \cQ$ and suppose
$P_{\al}, Q_{\al}\commutes E$.
Then the following relations hold.
\benum
\item[{\rm (i)}]
 $\displaystyle
  \left(\Inf_{\al} f(P_{\al},Q_{\al})\right)\And E
=
\left(\Inf_{\al} f(P_{\al}\And E,Q_{\al}\And E)\right)\And E$.
\item[{\rm (ii)}]
$\displaystyle
\left(\Sup_{\al} f(P_{\al},Q_{\al})\right)\And E
=
\left(\Sup_{\al} f(P_{\al}\And E,Q_{\al}\And E)\right)\And E$.
\eenum
\eTheorem 
\bProof
By assumption we have $\{P_\al,Q_\al\}^{!!}\subseteq\{E\}^{!}$.
It follows from (L1) that $f(P_\al,Q_\al)\in\{P_\al,Q_\al\}^{!!}$
 so that $f(P_\al,Q_\al)
\commutes E$.
From Proposition \ref{th:logic} and (L2) we have
\begin{align*}
  \left(\Sup_{\al} f(P_{\al},Q_{\al})\right)\And E
&=
\Sup_{\al}   \left(f(P_{\al},Q_{\al})\And E\right)\\
&=
\Sup_{\al}   \left(f(P_{\al}\And E,Q_{\al}\And E)\And E\right).
\end{align*}
Since $P_{\al},Q_{\al}\in\{ E\}^{!}$, we have $P_{\al}\And E,Q_{\al}\And E\in\{ E\}^{!}$,
and hence $\{P_{\al}\And E,Q_{\al}\And E\}^{!!}\subseteq\{ E\}^{!}$.
By (L1) we have $f(P_{\al}\And E,Q_{\al}\And E)\in \{P_{\al}\And E,Q_{\al}\And E\}^{!!}
\subseteq\{ E\}^{!}$, so that $f(P_{\al}\And E,Q_{\al}\And E)\commutes E$.
From Proposition \ref{th:logic} and (L2) we have
\begin{align*}
\Sup_{\al}   \left(f(P_{\al}\And E,Q_{\al}\And E)\And E\right)
&=
\left(\Sup_{\al}   f(P_{\al}\And E,Q_{\al}\And E)\right)\And E.
\end{align*}
Thus, relation (ii) follows.
Relation (i) follows similarly.
\eProof

\sloppy
The following theorem provides an important property of
ortholattice polynomials \cite[Proposition 3.1]{17A2}.
\bTheorem 
\label{th:restriction_property}
Every two-variable ortholattice polynomial on a logic $\cQ$ is a local
binary operation.
\eTheorem 

\subsection{Quantizations of classical connectives}
In this section we introduce a new method for studying the properties 
of ortholattice polynomials, using a simple application of the Decomposition Theorem.

Let $P,Q\in \cQ$. By Theorem \ref{th:decomposition}, 
the sublogic $\{P,Q\}^{!!}$ generated by $P,Q$ can be factored into 
the complete Boolean algebra $[0,\cdom(P,Q)]_{\{P,Q\}^{!!}}$
and the complete orthomodular lattice 
$[0,\cdom(P,Q)^{\perp}]_{\{P,Q\}^{!!}}$ without non-trivial
Boolean factor, where 
\beql{noncomM}
\cdom(P,Q)\p=(P\Or Q)\And (P\Or Q\p)\And(P\p\Or Q)\And(P\p\Or Q\p),
\eeq
from \Eq{comM}.  For any $X\in\{P,Q\}^{!!}$, define $X_B$ and $X_N$ by
\deq{X_B&=X\And \cdom(P,Q),\\ X_N&=X\And\cdom(P,Q)\p.}
Then any $X\in\{P,Q\}^{!!}$
can be uniquely decomposed as $X=X_B\Or X_N$ with the
condition that $X_B\le\cdom(P,Q)$ and $X_N\le\cdom(P,Q)\p$.
By \Eq{comM} and \Eq{noncomM}, we have $P^{\si}\And Q^{\ta}\le\cdom(P,Q)$
and $\cdom(P,Q)\p\le P^{\si}\Or Q^{\ta}$, where $\si,\ta\in\{\id,\perp\}$.
Thus, we have
\begin{align}
(P^{\si}\And Q^{\ta})_B
&=P^{\si}\And Q^{\ta},\label{eq:B-1}\\ 
(P^{\si}\And Q^{\ta})_N
&=0,\label{eq:N-1}\\
(P^{\si}\Or Q^{\ta})_B
&=\Sup_{\si':\si'\ne\si;\ta':\ta'\ne\ta} (P^{\si'}\And Q^{\ta'}),
\label{eq:B-2}\\ 
(P^{\si}\Or Q^{\ta})_N
&=\cdom(P,Q)\p.\label{eq:N-2}
\end{align}

\sloppy
A logic $\cQ$ is said to be {\em totally noncommutative} iff 
$\cdom(\cQ)=0$,
and {\em extremely noncommutative} iff 
\[
\Sup\{\cdom(P,Q)\mid  Q\not\in \{P, P\p\}
\mb{ and }P,Q\in \cQ\setminus \{0,1\}\}=0.
\]
\bProposition
A logic $\cQ$ is extremely noncommutative if and only if
 $P\And Q=0$ for any $P,Q\in\cQ\setminus\{1\}$ with $P\ne Q$.
\eProposition
\bProof
Suppose $\cQ$ is extremely noncommutative.  
Let $P,Q\in\cQ\setminus\{1\}$ with $P\ne Q$.
If $P=0$, $Q=0$, or $P=Q\p$, then $P\And Q=0$,
and otherwise $\cdom(P,Q)=0$ by assumption, so that 
$P\And Q\le \cdom(P,Q)=0$.  
Conversely suppose that $P\And Q=0$ for any 
$P,Q\in\cQ\setminus\{1\}$ with $P\ne Q$.
Suppose $0<P,Q<1,\ Q\not\in \{P, P\p\}$.
Then, $P\ne Q$, $P\ne Q\p$, $P\p\ne Q$, $P\p\ne Q\p$,
so that $P\And Q=P\And Q\p=P\p\And Q=P\p\And Q\p=0$,
and hence $\cdom(P,Q)=0$.  Thus, $\cQ$ is extremely noncommutative.
\eProof

Two examples of extremely noncommutative logic are in order:
(i) The modular lattice  MO2=$\{0,P,P\p,Q,Q\p,1\}$ 
 called the Chinese Lantern \cite[p.~16]{Kal83}. (ii) The projection lattice $\cQ(\C^2)$
of the 2-dimensional Hilbert space $\C^2$.

\sloppy
We obtain the following characterization of the two-variable ortholattice polynomials 
on a logic, originally obtained by Kotas \cite{Kot67}, as a straightforward consequence of
the Decomposition Theorem (Theorem \ref{th:decomposition}).

\bTheorem 
\label{th:poly-1}
Two-variable ortholattice polynomials $p(P,Q)$ in $P,Q$ over a logic 
$\cQ$ have the following form.
\beql{Kotas}
p(P,Q)=(P\And Q\And \al)\Or (P\And Q\p\And \be)\Or(P\p\And Q\And \ga)\Or(P\p\And Q\p
\And \de)
\Or (\ep\And \cdom(P,Q)\p),
\eeq
where $\al,\be,\ga,\de\in\{0,1\}$ and $\ep\in\{0,P,P\p,Q,Q\p,1\}$.
They define all the 16 Boolean operations 
\deq{
p(P,Q)=(P\And Q\And \al)\Or (P\And Q\p\And \be)\Or(P\p\And Q\And \ga)\Or(P\p\And Q\p
\And \de)
}
for $\al,\be,\ga,\de\in\{0,1\}$ on $\cQ$ if $\cQ$ is Boolean, \ie,
$\cdom(\cQ)=1$.
They define exactly 6 different monomials 
\deq{
p(P,Q)=\ep,
}
for $\ep\in\{0,P,P\p,Q,Q\p,1\}$ on $\cQ$ if $\cQ$ is extremely noncommutative.
They define 96 different operations on $\cQ$ 
if $\cQ$ is not Boolean nor extremely noncommutative.
\eTheorem
\bProof 
Let $p(P,Q)$ be an ortholattice polynomial in $P,Q$. 
Since $p(P,Q)\in \{P,Q\}^{!!}$, we have $p(P,Q)=p(P,Q)_B\Or p(P,Q)_N$.
Let $q(P,Q)$ be thedisjunctive normal form of $p(P,Q)$.
Then,  $p(P,Q)\And\cdom(P,Q)=q(P,Q)\And\cdom(P,Q)$ by the distributive law
and De Morgan's Laws for the Boolean algebra  $[0,\cdom(P,Q)]_{\{P,Q\}^{!!}}$,
and $q(P,Q)\And\cdom(P,Q)=q(P,Q)$ by \Eq{comM}.
Thus, we have $p(P,Q)_B=q(P,Q)$.
By De Morgan's Laws, we can assume that $p(P,Q)$ is a lattice polynomial in 
$P, P\p, Q, Q\p$ without any loss of generality.  Then, it follows from \Eq{N-1} and \Eq{N-2}
that
$
p(P,Q)_N=\ep\And\cdom(P,Q)^\perp,
$
where $\ep\in\{0,P,P\p,Q,Q\p,1\}$.
Thus, \Eq{Kotas} follows.  

If $\cQ$ is Boolean, we have $p(P,Q)=q(P,Q)$ for all
$P,Q\in\cQ$, so that $p(P,Q)$ defines at most 16 Boolean operations on $\cQ$.
In every $\cQ$ the Boolean subalgebra $\{0,1\}$ is invariant under any polynomials 
$p(P,Q)$, which define 16 different operations on $\{0,1\}$.  Thus, the polynomials
$p(P,Q)$ define all the 16 Boolean operations on $\cQ$ if $\cQ$ is Boolean.

Suppose that $\cQ$ is extremely noncommutative.
Let $P,Q\in\cQ$.  
If $P,Q\in\{0,1\}$ or $Q\in \{P, P\p\}$,  the value of $p(P,Q)$ is constant 
or dependent only on $P$ or $Q$.
Suppose  $0<P,Q<1$ and $Q\not\in \{P, P\p\}$.
Then, we have $\cdom(P,Q)=0$.
Hence, $p(P,Q)=\ep\And\cdom(P,Q)^\perp$ with  $\ep\in\{0,P,P\p,Q,Q\p,1\}$
defines at most 6 monomials on $\cQ$.  In this case, the set 
$\{0,P,P\p,Q,Q\p,1\}$ must have 6 different elements, 
otherwise we would have $\cdom(P,Q)=1$.
Thus, $p(P,Q)$ defines exactly 6 operations on $\cQ$.

Suppose that $\cQ$ is not Boolean nor extremely noncommutative.
In this case, there exists a pair $P,Q\in\cQ$ such that 
$P\not\in\{Q,Q\p\}$ and $0<E=\cdom(P,Q)<1$;
in fact,  since $\cQ$ is not extremely noncommutative, if $P\not\in\{Q,Q\p\}$
then $0<\cdom(P,Q)$, and since $\cQ$ is not Boolean there exists a pair
$P,Q\in\cQ$ such that $P\not\in\{Q,Q\p\}$ and $0<\cdom(P,Q)<1$. 
 By the Decomposition Theorem, in this case, $\cR=\{P,Q\}^{!!}$ is 
 the direct product of the Boolean algebra $\cR_B$
isomorphic to $[0,\cdom(\cQ)]$ 
and the complete orthomodular lattice $\cR_N$ isomorphic to 
$[0,\cdom(\cQ)\p]$ such that $\cdom(\cR_N)\p=1_{\cR}$, where $1_{\cR}$ is
the unit of $\cR$.  According to the arguments already given above, 
$p(P,Q)$ defines16 different operations on $\cB$ and 
16 different operations on $\cR$.  Therefore, $p(P,Q)$ defines exactly 96 
($=16\times 6$) operations on $\cQ$.
\eProof

A local binary operation $f(P,Q)$ on $\cQ$ is called a
{\em quantization} of a Boolean polynomial $b(P,Q)$ iff
$f(P,Q)_B=b_n(P,Q)$ for all $P,Q\in\cQ$, where $b_n(P,Q)$ is the disjunctive
normal form of $b(P,Q)$, and moreover $f(P,Q)$ is called a {\em polynomial
quantization} of  $b(P,Q)$ iff $f(P,Q)$ is polynomially definable.

The following theorem holds.

\bProposition\label{th:Kotas-2}
Let $f(P,Q)$ be a local binary operation on a logic $\cQ$
and $b(P,Q)$ a Boolean polynomial. 
The following statements are mutually equivalent.
\benum
\item[{\rm (i)}] $f(P,Q)$ is a quantization of $b(P,Q)$.
\item[{\rm (ii)}] If $P\commutes Q$, then $f(P,Q)=b(P,Q)$ for all $P,Q\in\cQ$.
\eenum
\eProposition
\bProof
(i)$\THEN$(ii):  Let $f(P,Q)$ be a quantization of $b(P,Q)$, \ie, 
$f(P,Q)_B=b_n(P,Q)$.
Suppose  $P\commutes Q$.  Then $\{P,Q\}^{!!}$ is Boolean
and $f(P,Q)\in\{P,Q\}^{!!}$ by (L1), so that $f(P,Q)=f(P,Q)_B
=b_n(P,Q)=b(P,Q)$. Thus, the assertion follows.

(ii)$\THEN$(i): Let $E=\cdom(P,Q)$.
Since $P,Q\commutes E$, from property (L2) 
we have
\[
f(P,Q)_B=f(P,Q)\And E=f(P\And E,Q\And E)\And E.
 \]
Since $P\And E\commutes Q\And E$, we have
$f(P\And E,Q\And E)=b(P\And E,Q\And E)$ by assumption.
Since the sublogic $\{P\And E,Q\And E\}^{!!}$
generated by $P\And E$ and $Q\And E$ is a Boolean sublogic, 
in which $b(P\And E,Q\And E)$ equals its disjunctive normal form 
$b_n(P\And E,Q\And E)$, \ie, $b(P\And E,Q\And E)=b_n(P\And E,Q\And E)$.
Thus, we have 
\begin{align*}
f(P\And E,Q\And E)\And E
&=b_n(P\And E,Q\And E)\And E.
%&=b_n(P,Q)\And E\\
%&=b_n(P,Q),
\end{align*}
From Theorem \ref{th:restriction_property} we have
$b_n(P\And E,Q\And E)\And E=b_n(P,Q)\And E$.
Since $b_n(P,Q)\le \cdom(P,Q)$ by \Eq{comM},
we have $b_n(P,Q)\And E=b_n(P,Q)$.
Thus, we have 
\[f(P,Q)_B=b_n(P,Q)
\]
and assertion (i) follows.
 \eProof
 
It follows from Theorem \ref{th:poly-1} that 
for each two-variable 
Boolean-polynomial $b(P,Q)$ there are exactly 6 polynomial quantizations
 $p(P,Q)$ of  $b(P,Q)$, which satisfy
\beql{quantization}
p(P,Q)=b_n(P,Q)\Or (\ep\And \cdom(P,Q)\p),
\eeq
where $b_n(P,Q)$ is the disjunctive normal form of $b(P,Q)$ and
$\ep\in\{0,P,P\p,Q,Q\p,1\}$.

\subsection{Quantizations of implication}
\label{se:PIIQL}

In classical logic, the implication connective 
$\Then$ is defined using negation $\perp$ and
disjunction $\Or$ by $P\Then Q=P^{\perp}\Or Q$.
In quantum logic, several counterparts have been proposed.  
Hardegree \cite{Har81} proposed the following
requirements, as ``minimal implicative conditions'', for the implication connective $\Then$.
\benum
\item[(LB)] If $P\commutes Q$, then 
$P\Then Q=P^{\perp}\Or Q$ for all $P,Q\in\cQ$.
\item[(E)]    $P\Then Q=1$ if and only if $P\le Q$ for all $P,Q\in\cQ$.
\item[(MP)] ({\it modus ponens}) $P             \And (P\Then Q) \le Q$ for all $P,Q\in\cQ$.
\item[(MT)] ({\it modus tollens}) $Q^{\perp}\And (P\Then Q) \le P^{\perp}$ for all $P,Q\in\cQ
$.
\item[(NG)] $P\And Q\p\le(P\Then Q)^\perp$ for all $P,Q\in\cQ$.
\end{enumerate}

A local binary operation $\Then$ on a logic $\cQ$ is called a {\em quantized implication} iff it 
is a quantization of classical implication $b(P,Q)=P\p\Or Q$,
or, equivalently, it satisfies (LB) by Proposition \ref{th:Kotas-2}.
A quantized implication $P\Then Q$ on $\cQ$ is called a 
{\em polynomially quantized implication} 
or said to be {\em polynomially definable} iff there exists a two-variable 
ortholattice polynomial $p(P,Q)$ in $P,Q$ such that $p(P,Q)=P \Then Q$ 
for all $P,Q\in\cQ$.  The Kotas theorem (Theorem \ref{th:poly-1}) concludes.

\bTheorem\label{th:Kotas-3}
{There exist exactly 6 two-variable ortholattice polynomials $P\Then_j Q$ for $j=0,
\ldots,5$
satisfying (LB), given as follows.}
\benum
\item[{\rm (0)}] 
$P\Then_0 Q=b_n(P,Q)$.
\item[{\rm (1)}] 
$P\Then_1 Q=b_n(P,Q)\Or (P\And \cdom(P,Q)\p)$.
\item[{\rm (2)}] 
$P\Then_2 Q=b_n(P,Q)\Or (Q\And \cdom(P,Q)\p)$.
\item[{\rm (3)}] 
$P\Then_3 Q=b_n(P,Q)\Or (P\p\And \cdom(P,Q)\p)$.
\item[{\rm (4)}] 
$P\Then_4 Q=b_n(P,Q)\Or (Q\p \And \cdom(P,Q)\p)$.
\item[{\rm (5)}] 
$P\Then_5 Q=b_n(P,Q)\Or \cdom(P,Q)\p$.
\end{enumerate}
In the above, $b_n(P,Q)$ is the disjunctive normal form of $b(P,Q)=P\p\Or Q$, \ie
\[
b_n(P,Q)=(P\p\And Q\p)\Or(P\p\And Q)\Or(P\And Q).
\]
\eTheorem

For $j=0,\ldots,5$, the above polynomials $P\Then_j Q$ are explicitly expressed as follows.
\benum\item[{\rm (0)}] 
 $P\Then_0 Q=(P\p\And Q\p)\Or(P\p\And Q)\Or(P\And Q)$.
\item[{\rm (1)}] 
 $P\Then_1 Q=(P\p\And Q\p)\Or(P\p \And Q)\Or(P\And(P\p\Or Q))$.
\item[{\rm (2)}] 
 $P\Then_2 Q=(P\p\And Q\p)\Or Q$.
\item[{\rm (3)}] 
 $P\Then_3 Q=P\p \Or(P\And Q)$.
\item[{\rm (4)}] 
 $P\Then_4 Q=((P\p\Or Q)\And Q\p)\Or(P\p \And Q)\Or(P\And Q)$.
\item[{\rm (5)}] 
 $P\Then_5 Q=P\p\Or Q$.
\end{enumerate}

The following characterizations of quantized implications hold
 \cite[Proposition 3.2]{17A2}.
\bProposition \label{th:implications-1}
Let $\Then$ be a local binary operation on a logic $\cQ$.
Then the following conditions are equivalent.
\benum\item[{\rm (i)}] $\Then$ is a quantized implication, i.e., it satisfies (LB).
\item[{\rm (ii)}] $(P\Then Q)_B=P\Then_0 Q$ for all $P,Q\in\cQ$.
\item[{\rm (iii)}] $(P\Then Q) \Or\cdom(P,Q)\p = P\Then_5Q$ for all
$P,Q\in\cQ$.
\item[{\rm (iv)}] $P\Then_0Q \le P\Then Q\le P\Then_5 Q$ for all  $P,Q\in\cQ$.
\end{enumerate}
\eProposition 

Note that every quantized implication $\Then$ has the property that 
$P\Then Q=1$ if $P\le Q$, since if   $P\le Q$, then $P\commutes Q$,
so that $P\Then Q=P\p\Or Q\ge P\p\Or P=1$.

In classical logic, condition (E) uniquely determines $\Then=P\p\Or Q$ up to
Boolean equivalence.  
In quantum logic, (E) implies (LB), whereas 
$P\Then_5 Q=P\p\Or Q$ satisfies (LB) but
does not satisfy (E), as shown in what follows.

\bTheorem\label{th:poly-2-1}
A two-variable ortholattice polynomial $P\Then Q$ satisfies (E) 
if and only if it satisfies (LB) and $(P\Then Q)_N\in\{0_N,P_N,P\p_N,Q_N,Q\p_N\}$.
\eTheorem
\bProof 
(only if part): 
Suppose that $P\Then Q$ satisfies (E). 
Suppose $P\commutes Q$.  Then $\{P,Q\}^{!!}$ is a Boolean algebra. 
By the truth table argument, (E) implies $P\Then Q=P\p\Or Q$.  Thus, (LB) holds.
From Theorem \ref{th:poly-1}, for general $P,Q\in\cQ$ we have
$$
P\Then Q=(P\And Q)\Or (P^{\perp}\And Q)\Or(Q^{\perp}\And P)
\Or (\ep\And \cdom(P,Q)\p),
$$
where $\ep\in\{0,P,P\p,Q,Q\p,1\}$.
Suppose $\ep=1$, \ie, $(P\Then Q)_N=\cdom(P,Q)\p$.
In $\cQ$=MO2, for instance, there exist $P,Q\in\cQ$ with $\cdom(P,Q)=0$, 
for which $P\Then Q=1$ holds but $P\le Q$ does not hold. 
This contradicts (E).  
Thus, (E) implies $(P\Then Q)_N\in\{0,P_N,P\p_N,Q_N,Q\p_N\}$.

\sloppy
(if part): Conversely, suppose that $\Then$ satisfies (LB)
and $(P\Then Q)_N\in\{0_N,P_N,P\p_N,Q_N,Q\p_N\}$.
If $P\le Q$, then $P\commutes Q$ and $P\Then Q=P\p\Or Q=1$,
so that $P\le Q$ implies $P\Then Q=1$.
Thus, it suffices to show that $P\Then Q=1$ implies $P\le Q$. 
Suppose $P\Then Q=1$.
Then $(P\Then Q)_B=\cdom(P,Q)$ and $(P\Then Q)_N=\cdom(P,Q)\p$.
Since $(P\Then Q)_B=(P\p\Or Q)_B$, it follows from $(P\Then Q)_B=\cdom(P,Q)$ 
that $P_B\le Q_B$.  
Thus, it suffices to show that if either 
$(P\Then Q)_N=0_N$, $= P_N$, $=P\p_N$, $=Q_N$, or $=Q\p_N$,
the relation $(P\Then Q)_N=\cdom(P,Q)\p$ implies $(P\Then Q)_N=0$.
If $(P\Then Q)_N=0$, this is obvious.
Suppose $(P\Then Q)_N=P_N$.
Since $(P\Then Q)_N=\cdom(P,Q)\p$, we have
$P\And \cdom(P,Q)\p=\cdom(P,Q)\p$, and hence
 $Q\And \cdom(P,Q)\p=Q\And P\And \cdom(P,Q)\p=0$
 and  $Q\p\And \cdom(P,Q)\p=Q\p\And P\And \cdom(P,Q)\p=0$,
 so that $\cdom(P,Q)\p=[Q\And \cdom(P,Q)\p]\Or [Q\p\And \cdom(P,Q)\p]=0$.
 Thus, if $(P\Then Q)_N=P_N$, then $\cdom(P,Q)=1$.
 Similarly, either $(P\Then Q)_N=P\p_N$, $(P\Then Q)_N=Q_N$, or 
 $(P\Then Q)_N=Q\p_N$ implies $\cdom(P,Q)=1$.
 It follows that  $P\Then Q=1$ implies  $P\le Q$.  Therefore,
$(P\Then Q)_B=(P\p\Or Q)_B$ and 
 $(P\Then Q)_N\in\{0_N,P_N,P\p_N,Q_N,Q\p_N\}$
 implies (E).
  \eProof

  As far as our knowledge extends, only an exhaustive proof has been known
  for the following fact \cite[Theorem 15.3]{Kal83}.  

\bCorollary\label{th:poly-2-1-c}
There are exactly 5 two-variable ortholattice polynomials $P\Then_j Q$
with $j=0,\ldots,4$ that satisfy (E), and yet $P\Then_5 Q=P\p\Or Q$ 
does not satisfy (E).
\eCorollary
\bProof 
Among all the two-variable ortholattice polynomials $P\Then_j Q$ with 
 $j=0,\ldots,5$ that satisfy (LB), 
 the condition $(P\Then_j Q)_N\in\{0_N,P_N,P\p_N,Q_N,Q\p_N\}$ 
 is satisfied only by $P\Then_j Q$ with $j=0,\ldots,4$.
 Thus, Theorem \ref{th:poly-2-1} concludes that 
 there are exactly 5 two-variable ortholattice polynomials $P\Then_j Q$
with $j=0,\ldots,4$ that satisfy (E), but that $P\Then_5 Q=P\p\Or Q$ 
does not satisfy (E).
 \eProof

Quantized implications satisfying (MP), (MT), and (NG) are characterized,
respectively, as follows.

\bProposition \label{th:quantum_implication}
Let $\Then$ be a quantized implication on a logic $\cQ$.
Then the following statements hold.
\benum\item[{\rm (i)}] $\Then$ satisfies (MP) if and only if
 $P\And (P\Then Q)_N=0$ for all $P,Q\in\cQ$.
 \item[{\rm (ii)}]
 $\Then$ satisfies (MT) if and only if
 $Q\p\And (P\Then Q)_N=0$ for all $P,Q\in\cQ$.
 \item[{\rm (iii)}]
  $\Then$ always satisfies (NG).
 \end{enumerate}
\eProposition 
\bProof
(i)  Suppose that (MP) holds.
Then we have $P\And(P\Then Q)\le P\And Q$ and hence
\[
P\And(P\Then Q)_N= P\And (P\Then Q)\And\cdom(P,Q)\p \le P\And Q\And \cdom(P,Q)\p=0.
\]
Thus $P\And(P\Then Q)_N=0$.
Conversely, suppose  $P\And(P\Then Q)_N=0$.
Then we have
\deqs{
P\And(P\Then Q) &=(P_B\And(P\Then Q)_B)\Or (P_N\And(P\Then Q)_N)\\
&=P_B\And (P\p\Or Q)_B\le Q_B\le Q.
}
Thus (MP) holds, and assertion (i) follows.

(ii) Suppose that (MT) holds.
Then we have  $Q\p\And (P\Then Q)\le  Q\p\And P\p$, and hence
 \[
 Q\p\And (P\Then Q)_N=Q\p\And (P\Then Q)\And \cdom(P,Q)\p
 \le Q\p\And P\p\And  \cdom(P,Q)\p=0.
 \]
 Thus $Q\p\And (P\Then Q)_N=0$. 
 Conversely, suppose $Q\p\And (P\Then Q)_N=0$ holds.
 We have
 \deqs{
Q\p\And(P\Then Q) &=(Q\p_B\And (P\Then Q)_B\Or (Q\p_N\And (P\Then Q)_N)\\
&=[Q\p\And (P\p\Or Q)]_B\le P\p_B\le P\p.
}
Thus (MT) holds, and assertion (ii) follows. 

(iii) From Theorem  \ref{th:Kotas-3} we have 
$P\Then Q\le b_n(P,Q)\Or\cdom(P,Q)\p \le P\p\Or Q$.  
Taking orthocomplements, assertion (iii) follows.
\eProof

As a result, the polynomially quantized implications satisfying (MP), (MT), and (NG) can be 
characterized, respectively, as follows.

\bTheorem\label{th:poly-2-2}
For any two-variable ortholattice polynomial $P\Then Q$ satisfying (LB), the following 
statements hold.
\begin{enumerate}[(R10)]
\item[{\rm (i)}]  $P\Then Q$ satisfies  (MP) if and only if 
$(P\Then Q)_N\in\{0_N,P\p_N,Q_N,Q\p_N\}$.
\item[{\rm (ii)}]  $P\Then Q$ satisfies  (MT) if and only if 
$(P\Then Q)_N\in\{0_N,P_N,P\p_N,Q_N\}$.
\item [{\rm (iii)}] $P\Then Q$ always satisfies  (NG).
\end{enumerate}
\eTheorem
\bProof 
The assertions follow easily from Proposition \ref{th:quantum_implication}.
\eProof
  
  Hardegree \cite[p.~189]{Har81} called a two-variable ortholattice polynomial 
  that satisfies all the minimum implicative conditions, 
    (E), (MP), (MT), and (NG), a
  {\em material implication}, and stated that there are exactly three
  material implications  $\Then_j$ with $j=0,2,3$,
  suggesting only an exhaustive proof.
  Here, we give an analytic proof for this statement.
 \bCorollary
\label{th:poly-3}
There are exactly three material implications: $\Then_0$, $\Then_2$, and $\Then_3$.
 \eCorollary
\bProof 
It follows from Theorems \ref{th:poly-2-1} and  \ref{th:poly-2-2} 
that a polynomially definable operation $P\Then Q$
satisfies (E), (MP), and (MT) if and only if 
\[
P\Then Q=(P\p\Or Q)_B\Or \ep_N
\]
for $\ep=\{0,P\p,Q\}$.
They correspond to $\Then_0$, $\Then_2$, and $\Then_3$.
\eProof

We call
$\Then_0$ the {\em minimum implication}, or {\em relevance implication} \cite{Geo79},
$\Then_2$ the {\em contrapositive Sasaki arrow},
$\Then_3$ the {\em Sasaki arrow}  \cite{Sas54,Fin69}, and 
$\Then_5$  the {\em classical implication}.  
So far we have no general
agreement on the choice from the above,  although the
majority view favors the Sasaki
arrow \cite{Urq83}.

\subsection{Quantizations of conjunction}
A local binary operation $\ast$  on a logic $\cQ$ is called a {\em quantized conjunction} iff it is 
a quantization of the classical conjunction
$b(P,Q)=b_n(P,Q)=P\And Q$, or equivalently, by Proposition \ref{th:Kotas-2},
the following condition is satisfied.
\benum\item[{\rm (GC)}] If $P\commutes Q$, then $P\ast Q=P\And Q$.
\end{enumerate}

In Boolean logic, implication and conjunction are associated by
the relation $P\And Q=(P\Then Q^{\perp})\p$, and this relation plays
an essential role in the duality between bounded universal quantification
$(\forall x\in u)\ph(x)$ and bounded existential quantification
$(\exists x\in u)\ph(x)$.  
In quantum logic, the truth value of the bounded universal quantification depends on
the choice of implication $\Then$ as
\[
\val{(\forall x\in u)\ph(x)}=\Inf_{x\in\dom(u)}(u(x)\Then \val{\ph(x)}).
\]
In order to maintain the duality, the bounded existential quantification 
should be defined as
\begin{align*}
\val{(\exists x\in u)\ph(x)}
&=\val{\Not (\forall x\in u)\Not \ph(x)}\\
&=\left(\Inf_{x\in\dom(u)}(u(x)\Then \val{\ph(x)}\p)\right)\p\\
&=\Sup_{x\in\dom(u)}(u(x)\Then \val{\ph(x)}\p)\p\\
&=\Sup_{x\in\dom(u)}(u(x)*\val{\ph(x)}),
\end{align*}
where $*$ is defined by 
\beql{DC}
P*Q=(P\Then Q\p)\p
\eeq
for all $P,Q\in\cQ$.
We call the operation $*$ defined in \Eq{DC} the {\em dual conjunction} 
of the quantized implication $\Then$.

For any $j=0,\ldots,5$ denote by $\ast_j$ the dual conjunction of the polynomial
implication $\Then_j$. Then
\benum
\item[{\rm (0)}] 
$P\ast_0 Q=(P\And Q)\Or \cdom(P,Q)\p$.
\item[{\rm (1)}] 
$P\ast_1 Q=(P\And Q)\Or (P\p\And \cdom(P,Q)\p)$.
\item[{\rm (2)}] 
$P\ast_2 Q=(P\And Q)\Or (Q\And \cdom(P,Q)\p)$.
\item[{\rm (3)}] 
$P\ast_3 Q=(P\And Q)\Or (P\And \cdom(P,Q)\p)$.
\item[{\rm (4)}] 
$P\ast_4 Q=(P\And Q)\Or (Q\p \cdom(P,Q)\p)$.
\item[{\rm (5)}] 
$P\ast_5 Q=P\And Q$.
\eenum

We call $\ast_5$ the {\em classical conjunction}, and $\ast_3$ the {\em Sasaki conjunction}.
If the implication $\Then$ is the classical one, \ie, $P\Then Q=
P\Then_5 Q=P\p\Or Q$, the dual conjunction $\ast_5$ is also the 
classical one, \ie,
$P\ast_5 Q=P \And Q$.  However, it is only in this case that
the classical conjunction appears, \eg,
the dual conjunction of the Sasaki arrow, 
$P\Then_3 Q=P\p\Or(P\And Q)$,
turns out to be the so called Sasaki projection, 
$P\ast_3 Q=P\And (P\p\Or Q)$
\cite{Sas54,Fin69}.
Some properties of $\ast_j$ for $j=0,\ldots,5$ were previously
studied by D'Hooghe and Pykacz \cite{HP00}.

We have the following.
\bProposition\label{th:dual_conjunction}
A binary operation $*$ on a logic $\cQ$ is a quantized conjunction if and only if
it is the dual conjunction of a quantized implication $\Then$ on $\cQ$.
\eProposition
\bProof
Let $\ast$ be the dual conjunction of a quantized implication $\Then$ on $\cQ$.
Since $\Then$ is local, we have 
$P\ast Q=(P\Then Q^{\perp})^{\perp}\in\{P,Q\}^{!!}$ by property (L1).
By the repeated use of property (L2), we have
\beqas
(P * Q)\And E
&=&[(P\Then Q^{\perp})^{\perp}]\And E\\
&=&
[(P\Then Q^{\perp})\And E]^{\perp}\And E\\
&=&
\{[(P\And E)\Then(Q^{\perp}\And E)]\And E\}^{\perp}\And E\\
&=&
[(P\And E)\Then(Q^{\perp}\And E)]^{\perp}\And E\\
&=&
[(P\And E)\Then[(Q\And E)^{\perp}\And E]]^{\perp}\And E\\
&=&
[(P\And E)\Then(Q\And E)^{\perp}]\And E]^{\perp}\And E\\
&=&
[(P\And E)\Then(Q\And E)^{\perp}]^{\perp}\And E\\
&=&
[(P\And E)*(Q\And E)]\And E.
\eeqas
Thus the operation $\ast$ is a local binary operation.
Property (GC) of $\ast$ easily follows from property (LB) of $\Then$.
To show the converse part, let $*$ be a quantized conjunction.
Let $\Then$ be defined by $P\Then Q=(P*Q\p)\p$ for all $P,Q\in\cQ$.
Then
$P\Then Q=(P*Q\p)\p\in\{P,Q\}^{!!}$, so that (L1) holds.
We have
\begin{align*}
(P\Then Q)\And E
&=(P*Q\p)\p\And E\\
&=[(P*Q\p)\And E]\p\And E\\
&=[(P\And E)*(Q\p\And E)]\p\And E\\
&=\{(P\And E)*[(Q\And E)\p\And E]\}\p\And E\\
&=[\{(P\And E)*[(Q\And E)\p]\}\And E]^\perp\And E\\
&=\{(P\And E)*(Q\And E)\p\}\p\And E\\
&=[(P\And E)\Then(Q\And E)]\And E,
\end{align*}
and hence (L2) holds.  Thus, $\Then$ is a quantized implication.
Since $(P\Then Q\p)\p=(P*Q\p\p)\p\p=P*Q$, the operation $*$ is the dual conjunction 
of a quantized implication $\Then$.  This completes the proof.
\eProof

We obtain the following characterizations of quantized conjunctions.
\bProposition \label{th:conjunctions-1}
Let $\ast$ be a local binary operation on a logic $\cQ$.
Then the following conditions are equivalent.
\benum
\item[{\rm (i)}] $\ast$ is a quantized conjunction, {\textit i.e.}, it satisfies (GC).
\item[{\rm (ii)}] $(P\ast Q)_B=P\And Q$ \quad for all $P,Q\in\cQ$.
\item[{\rm (iii)}] $(P\ast Q) \Or\cdom(P,Q)\p = 
 (P\p\Or Q)\And(P\Or Q\p)\And(P\Or Q)$\quad for all
$P,Q\in\cQ$.
\item[{\rm (iv)}] $P\And Q \le P\ast Q\le (P\p\Or Q)\And(P\Or Q\p)\And(P\Or Q)$
 \quad for all  $P,Q\in\cQ$.
\end{enumerate}
In particular,  a quantized conjunction $\ast$ satisfies
\beql{GC-1}
P\And Q\le P*Q\le P\Or Q.
\eeq
\eProposition 
\bProof
Since, by Proposition \ref{th:dual_conjunction}, every quantized conjunction
is the dual conjunction of a quantized implication,
the assertion can be derived from Proposition \ref{th:implications-1}
 by duality; note that conditions (ii) and (iii) are the duals of 
 conditions (iii) and (ii), respectively, in Proposition \ref{th:implications-1}. 
 Here, we alternatively give a direct proof.

(i)$\THEN$ (ii): 
Suppose (GC) is satisfied.  Let $P,Q\in\cQ$.
Since $P_B\commutes Q_B$, we have $P_B\ast Q_B
=P_B\And Q_B$, and $(P_B\And Q_B)\And \cdom(P,Q)=(P\And Q)\And \cdom(P,Q)=
P\And Q$.
Thus, from (L2) we have
\beqas
(P\ast Q) \And\cdom(P,Q)=(P_B\ast Q_B)\And \cdom(P,Q)
 =P\And Q,
\eeqas
and hence (i)$\THEN$(ii) follows. 

(ii)$\THEN$(iii): 
Suppose (ii) holds.
Note that $(P\ast Q)\Or \cdom(P,Q)^\perp=(P\ast Q)_B \Or \cdom(P,Q)^\perp$.
By taking the join with $\cdom(P,Q)^\perp$ in both sides of relation (ii), we have
$(P\ast Q)_B \Or \cdom(P,Q)^\perp= (P\And Q)\Or \cdom(P,Q)^\perp$.
Since $ (P\And Q)\Or \cdom(P,Q)^\perp
= (P\p\Or Q)\And(P\Or Q\p)\And(P\Or Q)$ by a calculation,
we obtain (iii), and the implication (ii)$\THEN$(iii) follows.

(iii)$\THEN$(iv): 
Suppose (iii) holds.  
Then $P\ast Q\le  (P\p\Or Q)\And(P\Or Q\p)\And(P\Or Q)$.  
By taking the meet with
 $\cdom(P,Q)$ in both sides of (iii), we have
 $(P\ast Q) \And \cdom(P,Q)= (P\And Q)\And \cdom(P,Q)$.
 Since $ (P\And Q)\And \cdom(P,Q)=P\And Q$, we have
$P\And Q\le P\ast Q$.
Thus the implication (iii)$\THEN$(iv) follows.  

(iv)$\THEN$(i): 
Suppose (iv) holds.  If $P\commutes Q$, we have
$P\And Q \le P\ast Q\le P\And Q$, so that $P\ast Q=P\And Q$.
Thus the implication (iv)$\THEN$(i) follows, and the
proof is completed.

\Eq{GC-1} follows from the relation $(P\p\Or Q)\And(P\Or Q\p)\And(P\Or Q)\le P\Or Q$.
 \eProof

The following proposition collects some useful relations.

\bProposition\label{th:logic2}
Let $\cQ$ be a  logic with a quantized implication $\Then$ and a quantized
conjunction $\ast$, and let $P,Q,P_\al,Q_\al,E\in\cQ$.
If $P,Q,P_\al,Q_\al,\commutes E$, then we have the following relations.
\benum
\item[{\rm (i)}] $P^{\perp}\And E=(P\And E)^{\perp}\And E$.

\item[{\rm (ii)}] $(P\And Q)\And E=[(P\And E)\And(Q\And E)].$

\item[{\rm (iii)}] $(P\Or Q)\And E=[(P\And E)\Or(Q\And E)].$

\item[{\rm (iv)}] $(P\Then Q)\And E=[(P\And E)\Then(Q\And E)]\And E.$

\item[{\rm (v)}] $\left(\Inf_{\al} (P_\al\Then Q_\al)\right)\And E=
\Inf_{\al}\left( (P_\al\And E)\Then (Q_\al\And E)\right)\And E.$

\item[{\rm (vi)}]  $\left(\Sup_{\al} (P_\al\ast  Q_\al)\right)\And E=
\Sup_{\al} \left((P_\al\And E)*(Q_\al\And E)\right).$
\eenum

\eProposition
\bProof
(i): The relation follows from focusing on $E$ (cf.~Proposition \ref{th:distributivity}).

(ii): The relation follows from associativity.

(iii): The relation follows from focusing on $E$ (cf.~Proposition \ref{th:distributivity}).

(iv): The relation follows from the locality of $\Then$.

(v): The relation follows from the locality of $\Then$ with Theorem \ref{th:LO} (i).

(vi): The relation follows from the locality of $\ast$ with Theorem \ref{th:LO} (ii)
and the relation $[(P_\al \And E)*(Q_\al \And E)]\le E$ obtained from \Eq{GC-1}.

\eProof

\subsection{Polynomially indefinable operations}
\label{se:GIIQL}

Takeuti \cite{Ta81} first introduced a polynomially indefinable binary operation
in quantum logic, about which he wrote:
\begin{quote}
We believe that we have to study 
this type of new operation in order to see the whole picture of quantum 
set theory including its strange aspects. \cite[p.~303]{Ta81}
\end{quote} 
\noindent
In fact, Takeuti \cite{Ta81} introduced a binary operation $\circ_{\theta}$ on 
the logic $\cQ(\cH)$ of projections on a Hilbert space $\cH$ by 
\beqas
P\circ_{\theta}Q=Q+(e^{i\theta}-1)PQ+(e^{-i\theta}-1)QP+2(1-\cos\theta)PQP
\eeqas
for all $P,Q\in\cQ(\cH)$.  It is easily seen that 
\beqas 
P\circ_{\theta}Q=e^{i\theta P}Q e^{-i\theta P}
\eeqas
for all $P,Q\in\cQ(\cH)$.
If $P\commutes Q$, then  $P\circ_{\theta}Q=Q$.
The binary operation $f(P,Q)=P\circ_{\theta} Q$ is local, \ie, (L1) and (L2) hold.
However, it is not in general definable as an ortholattice polynomial, since $f(P,Q)$ is
not generally in $\Ga\{P,Q\}$ \cite[Proposition 4.2]{17A2}.

Examples of polynomially indefinable quantized implications $\Then$, 
which even satisfy (MP), have been derived from Takeuti's polynomially 
indefinable operation $\circ_{\theta}$ \cite{17A2}.
Those operations $\Then$ satisfy (L1), \ie, $P\Then Q \in\{P,Q\}^{!!}$, 
but do not satisfy the condition 
$P\Then Q \in\Ga_0\{P,Q\}$, which all the polynomial implications
satisfy; see \S 4 in Ref.~\cite{17A2} for an extensive account of polynomially
indefinable quantized implications.

Examples of polynomially indefinable quantized conjunctions $\ast$
are given in the following.
For $j=0,\ldots,5$, for a real parameter $\theta\in [0,2\pi)$, and for $i=0,1$,
we define new binary operations $\ast_{j,\theta,i}$
on $\cQ=\cQ(\cH)$ by 
\beqas
P\ast_{j,\theta,0}Q&=&P\ast_{j}(P\circ_{\theta} Q),\\
P\ast_{j,\theta,1}Q&=&(Q\p\circ_{\theta}P)\ast_{j} Q
\eeqas
for all $P,Q\in\cQ$.  Obviously, $\ast_{j,0,i} =\ast_{j}$ for $j=0,\ldots,5$
and $i=0,1$.
Then, we obtain the following relations (cf.~Proposition 4.1 in Ref.~\cite{17A2}).
\benum
\item[\rm (i)] $P\ast_{0,\theta,0} Q=P\ast_0 Q$.
\item[\rm (ii)] $P\ast_{1,\theta,0} Q=P\ast_1 Q$.
\item[\rm (iii)] $P\ast_{2,\theta,0} Q=(P\ast_0 Q)\Or ((P\circ_{\theta}Q)\And \cdom(P,Q)\p)$.
\item[\rm (iv)] $P\ast_{3,\theta,0} Q=P\ast_3 Q$.
\item[\rm (v)] $P\ast_{4,\theta,0} Q=(P\ast_0 Q)\Or ((P\circ_{\theta}Q\p)\And \cdom(P,Q)\p)$.
\item[\rm (vi)] $P\ast_{5,\theta,0} Q=P\ast_5 Q$.
\item[\rm (vii)] $P\ast_{0,\theta,1} Q=P\ast_0 Q$.
\item[\rm (viii)] $P\ast_{1,\theta,1} Q=(P\ast_0 Q)\Or ((Q\p\circ_{\theta}P\p)\And \cdom(P,Q)\p)
$.
\item[\rm (ix)] $P\ast_{2,\theta,1} Q=P\ast_2 Q$.
\item[\rm (x)] $P\ast_{3,\theta,1} Q=(P\ast_0 Q)\Or ((Q\p\circ_{\theta}P)\And \cdom(P,Q)\p)$.
\item[\rm (xi)] $P\ast_{4,\theta,1}Q=P\ast_4 Q$.
\item[\rm (xii)] $P\ast_{5,\theta,1} Q=P\ast_5 Q$.
\eenum

The following theorem shows 
the existence of quantized conjunctions that are not polynomially
definable.

\bTheorem\label{th:non-poly}
Quantized conjunctions $\ast_{1,\theta,1}$, $\ast_{2,\theta,0}$, 
$\ast_{3,\theta,1}$, and $\ast_{4,\theta,0}$ are not polynomially 
definable for any $\theta\in(0,2\pi)$.
\eTheorem
\bProof
By duality, the assertion follows immediately from Proposition 4.2 in Ref.~\cite{17A2}.
\eProof

\section{Quantum Set Theory}\label{se:4}
\subsection{Orthomodular-valued universe}
\label{se:UQ}

We denote by $V$ the universe of Zermelo--Fraenkel set theory
with the axiom of choice (ZFC).
Let $\cQ$ be a logic. 
For each ordinal $ {\al}$, let
\beq
V_{\al}^{(\cQ)} = \{u|\ u:\dom(u)\to \cQ \mbox{ and }
(\exists \be<\al)
\dom(u) \subseteq V_{\be}^{(\cQ)}\}.
\eeq
The {\em $\cQ$-valued universe} $\VQ$ is defined
by 
\beq
  \VQ= \bigcup _{{\al}{\in}\mbox{On}} V_{{\al}}^{(\cQ)},
\eeq
where $\mbox{On}$ is the class of all ordinals. 

In the case where $\cQ$ is a Boolean algebra, $\VQ$ reduces to
the Boolean-valued universe of set theory \cite{TZ73,Bel05}.

For every $u\in\VQ$, the rank of $u$, denoted by
$\rank(u)$,  is defined as the least $\al$ such that
$u\in \VQ_{\al+1}$.
It is easy to see that if $u\in\dom(v)$, then 
$\rank(u)<\rank(v)$.
An induction on the rank leads to the following \cite[p.~21]{Bel05}.
\bTheorem[Induction Principle for $\VQ$]
For any predicate $\ph(x)$,
\[
\forall u\in \VQ[\forall u'\in\dom(u)\ph(u')\Then\ph(u)]\Then \forall u\in\VQ\ph(u)
\]
\eTheorem

For $u\in\VQ$, we define the {\em support} 
of $u$, denoted by $L(u)$, by transfinite recursion on the 
rank of $u$, by the relation
\beq
L(u)=\bigcup_{x\in\dom(u)}L(x)\cup\{u(x)\mid x\in\dom(u)\}\cup\{0\}.
\eeq
For $\cA\subseteq\VQ$ we write 
$L(\cA)=\bigcup_{u\in\cA}L(u)$ and
for $u_1,\ldots,u_n\in\VQ$ we write 
$L(u_1,\ldots,u_n)=L(\{u_1,\ldots,u_n\})$.
Then we obtain the following characterization of
subuniverses of $\VQ$.

\bProposition\label{th:sublogic}
Let $\cR$ be a sublogic of a logic $\cQ$ and $\al$ an
ordinal. For any $u\in \VQ$, we have
$u\in\VL_{\al}$  if and only if
$u\in \VQ_{\al}$ and $L(u)\subseteq\cR$. 
In particular, $u\in\VL$ if and only if
$u\in\VQ$ and $L(u)\subseteq\cR$. 
Moreover,  for any $u\in\VL$, its rank in $\VL$ 
is the same as its rank in $\VQ$.
\eProposition
\bProof Immediate from transfinite induction on
$\al$.
\eProof

\subsection{Orthomodular-valued interpretations}
Let $\cL(\in)$ be the language of first-order theory with equality 
consisting of the negation symbol $\Not$, connectives $\And, \Or, 
\Implies$, binary relation symbols $=,\in$, bounded 
quantifier symbols $\forall x\in y$, $\exists x \in y$, unbounded
quantifier symbols $\forall x, \exists x$, and no constant symbols.
For any class $U$, the language $\cL(\in,U)$ is the one obtained 
by adding a name for each element of $U$.

To each statement $\ph$ of $\cL(\in,U)$, the satisfaction
relation
$\bracket{U,\in} \models \ph$ is defined by the following recursive rules:
\benum
\item[(i)]$ \bracket{U,\in} \models \Not \ph \iff  \bracket{U,\in}
\models
\ph 
\mbox{ does not hold}$. 
\item[(ii)] $ \bracket{U,\in} \models \ph_1 \And \ph_2 
\iff \bracket{U,\in}
\models \ph_1 
\mbox{ and } \bracket{U,\in} \models \ph_2$.
\item[(iii)] $ \bracket{U,\in} \models \ph_1 \Or \ph_2 
\iff \bracket{U,\in}
\models \ph_1 
\mbox{ or } \bracket{U,\in} \models \ph_2$.
\item[(iv)] $ \bracket{U,\in} \models \ph_1 \Then \ph_2 
\iff \mb{ if }\bracket{U,\in}\models \ph_1, 
\mbox{ then } \bracket{U,\in} \models \ph_2$.
\item[(v)] $ \bracket{U,\in} \models  (\forall x\in u)\,\ph(x) \iff
\bracket{U,\in} \models \ph(u') \mbox{ for all } u' \in u$.
\item[(vi)] $ \bracket{U,\in} \models  (\exists x\in u)\,\ph(x) \iff
\mbox{there exists $u' \in u$ such that $\bracket{U,\in} \models \ph(u')$.}$
\item[(vii)] $ \bracket{U,\in} \models  (\forall x)\,\ph(x) \iff
\bracket{U,\in} \models \ph(u) \mbox{ for all } u \in U.$
\item[(viii)] $ \bracket{U,\in} \models  (\exists x)\,\ph(x) \iff
\mbox{ there exists }u\in U\mbox{ such that }
\bracket{U,\in} \models \ph(u) $.
\item[(ix)] $ \bracket{U,\in} \models u = v
\iff u = v$. 
\item[(x)] $ \bracket{U,\in} \models u\in v
\iff u\in v$.
\end{enumerate}
Our assumption that $V$ satisfies ZFC
means that 
$\bracket{V,\in}\models \ph(u_1,\ldots,u_n)$ for 
any formula $\ph(x_1,\ldots,x_n)$ of $\cL(\in)$ 
provable in ZFC and for any $u_1,\ldots,u_n\in V$.

Denote by $\cS(\cQ)$ the set of statements in $\cL(\in,\VQ)$.
A {\em $\cQ$-valued interpretation} of $\cL(\in,\VQ)$ is a mapping 
$\cI{(\Then,\ast)}:\ph\in\cS(\cQ)\mapsto \valQ{\ph}\in\cQ$ determined with a pair 
$(\Then,\ast)$ of local binary operations on $\cQ$ by 
the following rules, (R1)--(R10),
recursive on the rank of the elements of $\VQ$ and the complexity of the formulas.

\benum
\item[(R1)]  
$ \valQ{\Not\ph} = \valQ{\ph}^{\perp}$.
\item[(R2)]  $ \valQ{\ph_1\And\ph_2} 
= \valQ{\ph_{1}} \And \valQ{\ph_{2}}$.
\item[(R3)]  $ \valQ{\ph_1\Or\ph_2} 
= \valQ{\ph_{1}} \Or \valQ{\ph_{2}}$.
\item[(R4)]  $ \valQ{\ph_1\rightarrow\ph_2} 
= \valQ{\ph_{1}} \Then \valQ{\ph_{2}}$.
\item[(R5)]  $ \valQ{(\forall x\in u)\, {\ph}(x)} 
= \Inf_{u'\in \dom(u)}
(u(u') \Then \valQ{\ph(u')})$.
\item[(R6)]  $ \valQ{(\exists x\in u)\, {\ph}(x)} 
= \Sup_{u'\in \dom(u)}
(u(u') \ast \valQ{\ph(u')})$.
\item[(R7)]  $ \valQ{(\forall x)\, {\ph}(x)} 
= \Inf_{u\in \VQ}\valQ{\ph(u)}$.
\item[(R8)]  $ \valQ{(\exists x)\, {\ph}(x)} 
= \Sup_{u\in \VQ}\valQ{\ph(u)}$.
\item[(R9)] $\valQ{u = v}=\valQ{(\forall x\in u)(x\in v)\And (\forall x\in v)(x\in u)}$.
\item[(R10)] $ \valQ{u \in v}=\valQ{(\exists x\in v)(x=u)}$.
\end{enumerate}

The following relations follow from the above rules.
\benum
\item[(A1)] $\valQ{u = v}
= \inf_{u' \in  \cD(u)}(u(u') \Then
\valQ{u'  \in v})
\And \inf_{v' \in   \cD(v)}(v(v') 
\Then \valQ{v'  \in u})$.
\item[(A2)] $ \valQ{u \in v} 
= \sup_{v' \in \cD(v)} (v(v')* \valQ{v' = u})$.
\eenum

For a sublogic $\cR$ of a logic $\cQ$ with a $\cQ$-valued interpretation $\cI(\Then,\ast)$,
we denote by $\val{\ph}_{\cR}$ the $\cR$-valued
truth value of a statement $\ph\in\cS(\cQ)$ 
determined by the $\cR$-valued interpretation $\cI(\Then_{\cR},\ast_{\cR})$,
where $\Then_{\cR}$ and $\ast_{\cR}$ are the restrictions 
of $\Then$ and $\ast$ to $\cR$, 
which are well-defined by the locality of $\Then$ and $\ast$. 

A formula in $\cL(\in)$ is called a {\em
$\De_{0}$-formula}  iff it has no unbounded quantifiers
$\forall x$ or $\exists x$.
The following theorem holds.

\sloppy
\bTheorem[$\De_{0}$-Absoluteness Principle]
\label{th:Absoluteness}
\sloppy  
Let $\cR$ be a sublogic of a logic $\cQ$ with a $\cQ$-valued interpretation 
$\cI(\Then,\ast)$ of $\cL(\in,\VQ)$.
For any $\De_{0}$-formula 
${\ph} (x_{1},{\ldots}, x_{n}) \in \cL(\in)$ and $u_{1},{\ldots}, u_{n}\in V^{(\cR)}$, 
we have
\[
\val{\ph(u_{1},\ldots,u_{n})}_{\cR}=
\val{\ph(u_{1},\ldots,u_{n})}_{\cQ}.
\]
\eTheorem
\bProof 
The assertion is proved by induction on the complexity
of the formulas and the ranks of the elements of $\VQ$.
Let $u, v\in\VL$.
By the induction hypothesis, for any $u'\in\dom(u)$ and $v'\in\dom(v)$
we have
$\valR{u'\in w}=\valQ{u'\in w}$,
$\valR{v'\in w}=\valQ{v'\in w}$,
and $\valR{v'=w}=\valQ{v'=w}$
for all $w\in \VQ$.
Thus 
\beqas
\valR{u=v}
&=&\Inf_{u'\in\dom(u)}(u(u') \Then\valR{u'\in v})
\And
\Inf_{v'\in\dom(v)}(v(v') \Then\valR{v'\in u})\\
&=&\Inf_{u'\in\dom(u)}(u(u') \Then\valQ{u'\in v})
\And
\Inf_{v'\in\dom(v)}(v(v')\Then \valQ{v'\in u})\\
&=&
\valQ{u=v},
\eeqas
and  we also have 
\beqas
\valR{u\in v}
&=&
\Sup_{v'\in\dom(v)}(v(v')*\valR{v'=u})\\
&=&
\Sup_{v'\in\dom(v)}(v(v')*\valQ{v'=u})\\
&=&
\valQ{u\in v}.
\eeqas
Thus the assertion holds for atomic formulas.
Any induction step adding a logical symbol works
easily, even when bounded quantifiers are concerned,
since the ranges of the supremum and the infimum 
are common for evaluating $\valR{\cdots}$ and 
$\valQ{\cdots}$. 
\eProof

Henceforth, for any $\De_{0}$-formula 
${\ph} (x_{1},{\ldots}, x_{n}) \in\cL(\in)$
and $u_1,\ldots,u_n\in\VQ$,
we abbreviate $\val{\ph(u_{1},\ldots,u_{n})}=
\valQ{\ph(u_{1},\ldots,u_{n})}$.

The universe $V$  can be embedded in
$\VQ$ by the following operation 
$\vee:v\mapsto\check{v}$ 
defined by the $\in$-recursion: 
for each $v\in V$, $\check{v} = \{\check{u}|\ u\in v\} 
\times \{1\}$.  For any $P\in\cQ$, define $\tP=\{\av{\ck{0},P}\}\in\VQ$.

\bProposition\label{th:interpretation}
In any $\cQ$-valued interpretation $\cI(\Then,\ast)$, the following relations hold.
\benum
\item[{\rm (i)}] $\val{u\in\ck{0}}=0$\ \  for any $u\in \VQ$.
\item[{\rm (ii)}]  $\val{\ck{0}=\ck{0}}=1$.
\item[{\rm (iii)}]  $\val{\ck{0}=\tP}=P\Then 0$\ \ for any $P\in \cQ$.
\item[{\rm (iv)}]  $\val{\ck{0}\in\tP}=P\ast 1$\ \  for any $P\in \cQ$.
\eenum
\eProposition
\bProof
Since $\dom(\ck{0})=\emptyset$,
relations (i) and (ii) follow from
\deqs{
\val{u\in\ck{0}}&= \Sup_{v\in\dom(\ck{0})}(\ck{0}(v)\ast\val{v=u})=0,\\
\val{\ck{0}=\ck{0}}&= \Inf_{u\in\dom(\ck{0})}(\ck{0}(u)\Then\val{u\in\ck{0}})
\And \Inf_{u\in\dom(\ck{0})}(\ck{0}(u)\Then\val{u\in\ck{0}})
=1.
}
Since $\dom(\tP)=\{\ck{0}\}$,
relations (iii) and (iv) follow from
\begin{align*}
\val{\ck{0}=\tP}
&=\Inf_{u\in\dom(\ck{0})}(\ck{0}(u)\Then \val{u\in\tP})\And
\Inf_{v\in\dom(\tP)}(\tP(v)\Then \val{v\in\ck{0}})\\
&=1\And (\tP(\ck{0})\Then 0)=P\Then 0.\\
\val{\ck{0}\in\tP}
&=\Sup_{u\in\dom(\tP)}(\tP(u)\ast \val{u=\ck{0}})\\
&=\tP(\ck{0})\ast\val{\ck{0}=\ck{0}}\\
&=P\ast 1.
\end{align*}
\eProof

\subsection{Transfer Principle: Necessity}
\label{se:ZFC}\label{se:TPQ}

In this section, we investigate the Transfer Principle, which gives any $\De_0$-formula 
provable in ZFC a lower bound for its truth value, which is determined by the degree 
of the commutativity of the elements of $\VQ$ appearing in the formula as constants.

Let $\cQ$ be a logic.  
Let $\cA\subseteq\VQ$.  The {\em commutator
of $\cA$}, denoted by $\cm(\cA)$, is defined by 
\beq
\cuniv(\cA)=\com (L(\cA)).
\eeq
For any $u_1,\ldots,u_n\in\VQ$, we write
$\cuniv(u_1,\ldots,u_n)=\cuniv(\{u_1,\ldots,u_n\})$.

Let $\cI(\Then,\ast)$ be a $\cQ$-valued interpretation.
We denote by $\val{\ph}$ the $\cQ$-valued
truth value of a statement $\ph\in\cS(\cQ)$ 
determined by the $\cQ$-valued interpretation $\cI(\Then,\ast)$.
Then, the Transfer Principle for the $\cQ$-valued interpretation $\cI(\Then,\ast)$
 is formulated as follows.
\medskip

{\bf Transfer Principle.}  
{\em Any $\De_{0}$-formula ${\ph} (x_{1},{\ldots}, x_{n})$ in
$\cL(\in)$ provable in ZFC satisfies
\beq
\val{\ph({u}_{1},\ldots,{u}_{n})}\ge
\cuniv(u_{1},\ldots,u_{n})
\eeq
}
 for any $u_1,\ldots,u_n\in\VQ$.
\medskip

A $\cQ$-valued interpretation $\cI(\Then,\ast)$ is called the {\em Takeuti interpretation}
iff $\Then$ is the Sasaki arrow and $\ast$ is the classical conjunction, \ie,
$P\Then Q=P\Then_3 Q=P\p\Or(P\And Q)$ and $P\ast Q=P\ast_5 Q=P\And Q$
for all $P,Q\in\cQ$.
It was shown that if $\cQ$ is the projection lattice of a von Neumann algebra,
then the $\cQ$-valued Takeuti interpretation $\cI(\Then_3,\ast_5)$ satisfies the
Transfer Principle \cite{07TPQ}.
This result was extended to an arbitrary logic $\cQ$ and arbitrary quantized
implication $\Then$ on $\cQ$ to show that any $\cQ$-valued interpretation
$\cI(\Then,\ast_5)$ satisfies the Transfer Principle \cite{17A2}.
In the present paper we consider the problem of finding all the interpretations that satisfy
the Transfer Principle.  

In order to eliminate uninteresting interpretations from our consideration,
we call a $\cQ$-valued interpretation $\cI(\Then, \ast)$ {\em non-trivial}
iff for any $P\in\cQ$ there exist a $\De_0$-formula $\ph(x_1,\ldots,x_n)\in\cL(\in)$
and $u_1,\ldots,u_n\in\VQ$ such that $\cm(u_1,\ldots,u_n)=1$ and 
$\val{\ph(u_1,\ldots,u_n)}=P$.
Simple sufficient conditions for non-triviality are given as follows.

\bProposition
If a $\cQ$-valued interpretation $\cI(\Then,\ast)$ satisfies
\benum
\item[{\rm (i)}] $P\Then 0=P\p$ \  for all $P\in\cQ$, or   

\item[{\rm (ii)}] $P\ast 1=P$ \   for all $P\in\cQ$, 
\eenum
then  $\cI(\Then,\ast)$ is non-trivial.
\eProposition
\bProof
Suppose condition (i) holds.
Let $\ph(x_1,x_2):=\Not(x_1=x_2)$, $u_1=\ck{0}$, and $u_2=\tP$.
Then $\cm(u_1,u_2)=1$ and $\val{\ph(u_1,u_2)}=\val{\ck{0}=\tP}\p=(P\Then 0)\p=P$
from Proposition \ref{th:interpretation} (iii).  Thus, the interpretation $\cI(\Then,\ast)$ is
non-trivial.
Suppose condition (ii) holds.
Let $\ph(x_1,x_2):=(x_1\in x_2)$,  $u_1=\ck{0}$, and $u_2=\tP$.
Then $\cm(u_1,u_2)=1$ and $\val{\ph(u_1,u_2)}=\val{\ck{0}\in \tP}=P\ast 1=P$
from Proposition \ref{th:interpretation} (iv).  Thus, the interpretation $\cI(\Then,\ast)$ is
non-trivial.
\eProof

In what follows, we introduce the connective $\IFF$ in the language
$\cL(\in\VQ)$ as an abbreviation for 
$\ph\IFF \ps:=(\ph\And \ps)\Or(\Not \ph\And \Not \ps)$
for any $\ph,\ps\in\cL(\in,\VQ)$
and the corresponding operation $\IFF$ on $\cQ$ by
$P\IFF Q:=(P\And Q)\Or(P\p \And Q\p)$ for all $P,Q\in\cQ$.

Then we have the following theorem, showing that in order 
for a non-trivial $\cQ$-valued interpretation $\cI(\Then,\ast)$
to satisfy the Transfer Principle, it is necessary that $\Then$ 
satisfies (LB) and that $\ast$ satisfies (GC).

\bTheorem\label{th:TP-1}
If a non-trivial $\cQ$-valued interpretation $\cI(\Then,\ast)$ of 
$\cL(\in,\VQ)$ satisfies the Transfer Principle,
then the operation $\Then$ is a quantized implication and the operation
$\ast$ is a quantized conjunction.
\eTheorem
\bProof
Let $P\in\cQ$.
By assumption, there exist a $\De_0$-formula $\ph(x_1,\ldots, x_n)\in\cLin$
and $u_1,\ldots ,u_n\in\VQ$ such that $\cm(u_1,\ldots, u_n)=1$
and $\val{\ph(u_1,\ldots, u_n)}=P$.
Since 
\[
[\ph(x_1,\ldots,x_n)\Then \Not(x_{n+1}=x_{n+1})]\IFF \Not\ph(x_1,\ldots,x_n)
\]
is provable in ZFC, by the Transfer Principle we have
\[
\val{[\ph(u_1,\ldots,u_n)\Then \Not(\ck{0}=\ck{0})]\IFF \Not\ph(u_1,\ldots,u_n)}
\ge\cm(u_1,\ldots,u_n,\ck{0}).
\]
Since $\dom(\ck{0})=\emptyset$, we have $\cm(u_1,\ldots,u_n,\ck{0})=
\cm(u_1,\ldots,u_n)=1$, so that we obtain
\[
\val{\ph(u_1,\ldots,u_n)}\Then\val{\ck{0}=\ck{0}}\p=\val{\ph(u_1,\ldots,u_n)}\p.
\]
Since $\val{\ck{0}=\ck{0}}=1$ by Proposition \ref{th:interpretation} (ii),
we have $P\Then 0=P\p$ for all $P\in\cQ$. 
Recall 
$\tP=\{\bracket{\ck{0},P}\}\in\VQ$.
Since $\cm(\tP,\ck{0})=1$, from the Transfer Principle
we obtain 
\deqs{
\val{(\exists x\in \tP)(x=\ck{0})\IFF
\Not(\forall x\in \tP)\Not(x=\ck{0})}=1.
}
Since $\dom(\tP)=\{\ck{0}\}$, we obtain
\deqs{
\val{\ck{0}\in \tP}
&=\val{(\exists x\in \tP)(x=\ck{0})}\\
&=\val{\Not(\forall x\in \tP)\Not(x=\ck{0})}\\
&=\left(\Inf_{u'\in \dom(\tP)}(\tP(u')\Then\val{u'=\ck{0}}\p)\right)\p\\
&=(\tP(\ck{0})\Then\val{\ck{0}=\ck{0}}\p)\p\\
&=(P\Then 0)\p.
}
Since  $P\Then 0=P\p$,  we have $\val{\ck{0}\in \tP}=P$.
Let $\vp(x_1,x_2,x_3)$ be the $\De_0$-formula in $\cL(\in)$ such that
$$
\vp(x_1,x_2,x_3):=(x_1\in x_2\Then x_1\in x_3)
\IFF \left(\Not(x_1\in x_2)\Or(x_1\in x_3)\right).
$$
Then  $\ZFC\vdash \vp(x_1,x_2,x_3)$.
Let $P,Q\in\cQ$ with $P\commutes Q$.
We have $\cm(\ck{0},\tP,\tQ)=\cdom(P,Q)=1$.  
By the Transfer Principle, we have 
$\val{\vp(\ck{0},\tP,\tQ)}\ge\cm(\ck{0},\tP,\tQ)=1$.
Thus, we have 
\beqas
\val{\ck{0}\in \tP}\Then \val{\ck{0}\in\tQ}=\val{\ck{0}\in\tP}\p\Or\val{(\ck{0}\in\tQ)},
\eeqas
and hence we conclude 
 $$
 P\Then Q=P^{\perp}\Or Q.
 $$
 Since $P,Q\in\cQ$ are arbitrary elements with $P\commutes Q$,
the operation $\Then$ satisfies (LB), and hence it is a quantized implication.

By the definition of the interpretation $\cI(\Then,\ast)$,
we have relation (A2), so that
\begin{align*}
\val{\tP\in \tQ}
&=\Sup_{v\in \dom(\tQ)}(\tQ(v)\ast \val{v=\tP})\\
&=\tQ(\ck{0})\ast\val{\ck{0}=\tP}\\
&=Q\ast P\p.
\end{align*}
On the other hand, since
\[
\ph(x_1,x_2):=x_1\in x_2
\IFF \Not(\forall y_1\in x_2)\Not(y_1=x_1)
\]
is a $\De_0$-formula provable in ZFC and $\cm(\tP,\tQ)=1$,
by the Transfer Principle we have $\val{\ph(\tP,\tQ)}=1$,
so that
\begin{align*}
\val{\tP\in \tQ}
&=\val{\Not(\forall v\in \tQ)\Not(v=\tP)}\\
&=\left(\Inf_{v\in \dom(\tQ)}(\tQ(v)\Then\val{v=\tP}\p)\right)\p\\
&=(\tQ(\ck{0})\Then\val{\ck{0}=\tP}\p)\p\\
&=(Q\Then P)\p.
\end{align*}
Since $P\commutes Q$ and $\Then$ satisfies (LB), we have $(Q\Then P)\p=Q \And P\p$.
Thus $\val{\tP\in \tQ}=Q \And P\p$.
It follows that $Q\ast P\p=Q\And P\p$. Since $P,Q\in\cQ$ were arbitrary commuting
 elements,  condition (GC) holds for the operation $\ast$.  Thus, $\ast$ is a quantized
 conjunction. 
\end{proof}

A $\cQ$-valued interpretation $\cI(\Then,\ast)$ is called {\em normal} iff
$\Then$ is a quantized implication and $\ast$ is a quantized conjunction.
It is easy to see that all normal interpretations are non-trivial.
It follows from Theorem \ref{th:TP-1} that all the non-trivial $\cQ$-valued
interpretations satisfying the Transfer Principle are normal. 
 
 \subsection{Transfer Principle: Sufficiency} 
In what follows,
suppose that for any $\ph\in\cL(\in,\VQ)$ the truth value $\val{\ph}\in\cQ$
is assigned by a fixed but arbitrary normal $\cQ$-valued interpretation
$\cI(\Then,\ast)$.
In this section, we shall prove that all the normal interpretations admit the Transfer
Principle.

The following theorem is known as the fundamental theorem of 
Boolean-valued models of set theory.
\bTheorem\label{th:BTP-1}
If $\cQ$ is a Boolean logic, all the normal interpretations define
a unique $\cQ$-valued interpretation and satisfy the following statements.
\benum
\item[{\rm (i)}] $\val{(\exists x\in u)\ph(x)}
=\val{(\exists x)(x\in u\And \ph(u))}$ for every formula $\ph(x)$ in 
$\cL(\in,\VQ)$ with one free variable $x$ and $u\in\VQ$.
\item[{\rm (ii)}]  $\val{(\forall x\in u)\ph(x)}
=\val{(\forall x)[\Not(x\in u)\Or\ph(x)]}$
for every formula $\ph(x)$ in 
$\cL(\in,\VQ)$ with one free variable $x$ and $u\in\VQ$.
\item[{\rm (iii)}] $\val{\ph}=1$ for any statement in $\cL(\in,\VQ)$ provable in ZFC.
\eenum
\eTheorem
\bProof Let $\cQ$ be a Boolean logic.
Let $\cI(\Then, \ast)$ be a normal $\cQ$-valued interpretation of
$\cL(\in,\VQ)$. By  normality, we have
\deqs{
\val{(\exists x\in u)\ph(x)}=\Sup_{u'\in\dom(u)}(u(u')\And \val{\ph(u')}).
}
Then statement (i) follows from the relation 
\deqs{
\val{(\exists x)(x\in u\And\ph(x))}=\Sup_{u'\in\dom(u)}(u(u')\And \val{\ph(u')})
}
which is well-known for Boolean-valued models
\cite[Theorem 13.13]{TZ73}, \cite[Corollary 1.18]{Bel05}.
Statement (ii) follows from (i) by duality between 
$\Then$ and $\ast$ on Boolean algebras.  
Thus, the interpretation is uniquely determined by the part of 
the language without bounded quantifiers.
Hence statement (iii) follows from the fundamental theorem
of Boolean-valued models \cite[Theorems 13.12 and 14.25]{TZ73},
\cite[Theorem 1.33]{Bel05}.
\eProof

Denote by ${\bf 2}$ the sublogic ${\bf 2}=\{0,1\}$ in any logic $\cQ$.
We have the following.
\bTheorem[$\De_0$-Elementary Equivalence Principle]
\label{th:2.3.2}
\sloppy
Let ${\ph} (x_{1},{\ldots}, x_{n}) $ be a 
$\De_{0}$-for\-mu\-la  of $\cL(\in)$.
For any $u_{1},{\ldots}, u_{n}\in V$,
we have
$$
\bracket{V,\in}\models {\ph}(u_{1},{\ldots},u_{n})
\quad\mbox{if and only if}\quad
\val{\ph(\check{u}_{1},\ldots,\check{u}_{n})}=1.
$$
\eTheorem
\bProof
By induction it is easy to see that 
$
\bracket{V,\in}\models  {\ph}(u_{1},{\ldots},u_{n})
\mbox{ if and only if }
\val{\ph(\check{u}_{1},\ldots,\check{u}_{n})}_{\bf 2}=1
$
for any ${\ph} (x_{1},{\ldots}, x_{n})$ in $\cL(\in)$, 
and this is
equivalent  to $\val{\ph(\check{u}_{1},\ldots,\check{u}_{n})}=1$
for any $\De_{0}$-formula ${\ph} (x_{1},{\ldots}, x_{n})$ 
by the $\De_0$-absoluteness principle.
\eProof

The following proposition is useful in applications.

\bProposition\label{th:elementary}
If $\dom(u)\subseteq \dom(\ck{X})$ for some $X\in V$,
then $\val{x\in u}=u(x)$ for any $x\in \dom(u)$ in any
normal $\cQ$-valued interpretation $\cI(\Then,\ast)$.
\eProposition
\bProof
Let $x\in \dom(u)$.
Since $\dom(u)\subseteq \dom(\ck{X})$, there
is some $x'\in X$ such that $x=\ck{x'}$.
We have 
\deqs{
\val{x\in u}&=\Sup_{u'\in\dom(u)}u(u')\ast \val{u'=x}\\
&=\Sup_{u'\in u}u(\ck{u'})\ast \val{\ck{u'}=\ck{x'}}\\
&=\Sup_{u'\in u}u(\ck{u'})\And \val{\ck{u'}=\ck{x'}}\\
&=\Sup_{u'=x'}u(\ck{u'})\\
&=u(\ck{x'})\\
&=u(x).
}
Thus, the assertion follows.
\eProof

Let $\cA\subseteq\VQ$.  The {\em logic
generated by $\cA$}, denoted by $\cQ(\cA)$, is  defined by 
\beq
\cQ(\cA)=L(\cA)^{!!}.
\eeq
For $u_1,\ldots,u_n\in\VQ$, we write
$\cQ(u_1,\ldots,u_n)=\cQ(\{u_1,\ldots,u_n\})$.

The following theorem shows that the Transfer Principle partially holds
if $\cm(u_1,\ldots,u_n)=1$.

\bTheorem\label{th:BTP-2}
For any $u_1,\ldots,u_n\in\VQ$ with $\cm(u_1,\ldots,u_n)=1$,
every $\De_{0}$-formula ${\ph} (x_{1},{\ldots}, x_{n})$ in
$\cL(\in)$ provable in ZFC holds with the truth value 1, \ie,
\deqs{
  \val{\ph({u}_{1},\ldots,{u}_{n})}=1.
}
\eTheorem
\bProof
Since $\cm(u_1,\ldots,u_n)=1$, $\cQ(u_1,\ldots,u_n)$ is a 
Boolean algebra.  Let $\cB=\cQ(u_1,\ldots,u_n)$.
Applying Theorem \ref{th:BTP-1} (iii) to 
the $\cQ$-valued interpretation $\cI(\Then,\ast)$ restricted 
to $\VB$, we have $\val{\ph({u}_{1},\ldots,{u}_{n})}_{\cB}=1$.
By the  $\De_0$-Absoluteness Principle, we have
\deqs{
 \val{\ph({u}_{1},\ldots,{u}_{n})}=
  \val{\ph({u}_{1},\ldots,{u}_{n})}_{\cB}=1,
}
and the proof is completed.
\eProof

Let $u\in\VQ$ and $p\in\cQ$.
The {\em restriction} $u|_p$ of $u$ to $p$ is defined by
the following transfinite recursion:
\beqas
u|_p=\{\av{x|_p,u(x)\And p}\mid x\in\dom(u)\}\cup\{\av{u,0}\}.
\eeqas
The last term $\{\av{u,0}\}$ has no essential role but 
ensures that the function $u|_p:\dom(u|_p)\to \cQ$ is well-defined, \ie,
if $u|_p=v|_p$, then $u=v$ and $u(x)\And p=v(x)\And p$ for all $x\in\dom(u)=\dom(v)$.
Note that our definition of restriction is simpler than
the corresponding notion given by Takeuti \cite
{Ta81}.
We shall develop the theory of restriction 
along a different line.

\bProposition\label{th:L-restriction}
For any $\cA\subseteq \VQ$ and $p\in\cQ$, 
we have 
\beq
L(\{u|_p\mid u\in\cA\})=L(\cA)\And p.
\eeq
\eProposition
\bProof
By induction, it is  easy to show that
$
L(u|_p)=L(u)\And p,
$
and hence the assertion follows easily.
\eProof

\bProposition\label{th:range}
For any $\De_0$-formula $\ph(x_1,\ldots,x_n)$ in
$\cL(\in)$ and $u_1,\cdots,u_n\in\VQ$,
the following statements hold.
\benum
\item[{\rm (i)}]
$\val{\ph(u_1,\ldots,u_n)}\in\cQ(u_1,\ldots,u_n)$.

\item[{\rm (ii)}]
 If 
$p\in L(u_1,\ldots,u_n)^{!}$, then 
$p\commutes \val{\ph(u_1,\ldots,u_n)}$
and $p\commutes \val{\ph(u_1|_p,\ldots,u_n|_p)}$.
\eenum
\eProposition

\bProof
(i): Let $\cA=\{u_1,\ldots,u_n\}$.
Since $L(\cA)\subseteq\cQ(\cA)$, it follows from
Proposition \ref{th:sublogic} that $u_1,\ldots,u_n\in
V^{(\cQ(\cA))}$.
By the $\De_0$-absoluteness
principle, we have 
$\val{\ph(u_1,\ldots,u_n)}=
\val{\ph(u_1,\ldots,u_n)}{}_{\cQ(\cA)}\in \cQ(\cA)$.

(ii)  Let $u_{1},{\ldots}, u_{n}\in\VQ$.
If $p\in L(u_1,\ldots,u_n)^{!}$, then
$p\in \cQ(u_1,\ldots,u_n)^{!}$.  From 
(i),
$\val{\ph(u_1,\ldots,u_n)}\in\cQ(u_1,\ldots,u_n)$,
so that $p\commutes \val{\ph(u_1,\ldots,u_n)}$.
From Proposition \ref{th:L-restriction},
$L(u_1|_p,\ldots,u_n|_p)=L(u_1,\ldots,u_n)\And p$,
and hence $p\in L(u_1|_p,\ldots,u_n|_p)^{!}$, so that
$p\commutes \val{\ph(u_1|_p,\ldots,u_n|_p)}$.
\eProof

We define the binary relation $x_1\subseteq x_2$ by
$\forall x\in x_1(x\in x_2)$.
Then, by definition for  any $u,v\in\VQ$ we have
\beq
\val{u\subseteq v}=
\Inf_{u'\in\dom(u)}(u(u') \rightarrow\val{u'\in v}),
\eeq
and we have $\val{u=v}=\val{u\subseteq v}
\And\val{v\subseteq u}$.

\bProposition\label{th:restriction-atom}
For any $u,v\in\VQ$ and $p\in L(u,v)^{!}$,
the following relations hold.
\benum
\item[{\rm (i)}] $\val{u|_p\in v|_p}\And p=\val{u\in v}\And p$.

\item[{\rm (ii)}] $\val{u|_p\subseteq v|_p}\And p
=\val{u\subseteq v}\And p$.

\item[{\rm (iii)}] $\val{u|_p= v|_p}\And p =\val{u= v}\And p$
\eenum
\eProposition
\bProof
We prove the relations by induction on the rank of 
$u,v$.  If $\rank(u)=\rank(v)=0$, then $\dom(u)=\dom(v)
=\emptyset$, so that the relations trivially hold.
Let $u,v\in\VQ$ and $p\in L(u,v)^{!}$.
To prove (i),
let $v'\in\dom(v)$. 
Then, we have $p\commutes v(v')$ by the assumption on $p$.
By the induction hypothesis, we also have
$\val{u|_p=v'|_p}\And p=\val{v'=u}\And p$.
By Proposition \ref{th:range} (ii), we have 
$p\commutes \val{v'=u}$, so that
$v(v'), \val{v'=u}\in\{p\}^{!}$, 
and hence $v(v')*\val{v'=u}\in\{p\}^{!}$ by locality. 
From Proposition \ref{th:logic2} (vi)  we have
\begin{align*}
\val{u|_p\in v|_p}\And p
&=
\left(\Sup_{v'\in\dom(v|_p)}(v|_p(v')*\val{u|_p=v'})\right)\And p\\
&=
\left(\Sup_{v'\in\dom(v)}(v|_p(v'|_p)*\val{u|_p=v'|_p})\right)\And p \\
&=
\left(\Sup_{v'\in\dom(v)}[(v|_p(v')\And p)*\val{u|_p=v'|_p}]\right)\And p \\
&=
\Sup_{v'\in\dom(v)}[(v(v')\And p)*(\val{u|_p=v'|_p}\And p)].
\end{align*}
Thus,  by the induction hypothesis and Proposition \ref{th:logic2} (vi)
we  have
\begin{align*}
\val{u|_p\in v|_p}\And p
&=
\Sup_{v'\in\dom(v)}[(v(v')\And p)*(\val{v'=u}\And p)]\\
&=
\left(\Sup_{v'\in\dom(v)}(v(v')*\val{v'=u})\right)\And p\\
&=
\val{u\in v}\And p.
\end{align*}  
Thus, relation (i) has been proved.
To prove (ii), let $u'\in\dom(u)$.
Then we have $\val{u'|_p\in v|_p}\And p=\val{u'\in v}\And p$
by the induction hypothesis.
Thus, by Proposition \ref{th:logic2} (v) we have
\beqas
\val{u|_p\subseteq v|_p}\And p
&=&
\left(\Inf_{u'\in\dom(u|_p)}(u|_p(u')\Then\val{u'\in v|_p})\right)\And p\\
&=&
\left(\Inf_{u'\in\dom(u)}(u|_p(u'|_p)\Then\val{u'|_p\in v|_p})\right)\And p\\
&=&
\Inf_{u'\in\dom(u)}\left([(u|_p(u'|_p)\And p)\Then(\val{u'|_p\in v|_p}\And p)]\And p\right)\\
&=&
\Inf_{u'\in\dom(u)}
[(u(u')\And p)\Then(\val{u'\in v}\And p)]\And p.
\eeqas
We  have $p\commutes u(u')$ by
assumption on $p$, and $p\commutes\val{u'\in v}$
by Proposition \ref{th:range} (ii),
so that
$p\commutes u(u')\Then\val{u'\in v}$ and
$p\commutes (u(u')\And p)\Then(\val{u'\in v}\And p)$.
From Proposition \ref{th:logic2} (iv) we have
$$
[(u(u')\And p)\Then(\val{u'\in v}\And p)]\And p
=
(u(u')\Then\val{u'\in v})\And p.
$$
Thus, we have
\beqas
\val{u|_p\subseteq v|_p}\And p
&=&
\left(\Inf_{u'\in\dom(u)}(u(u')\And p)\Then(\val{u'\in v}\And p)\right)\And p\\
&=&
\Inf_{u'\in\dom(u)}[(u(u')\And p)\Then(\val{u'\in v}\And p)]\And p\\
&=&
\Inf_{u'\in\dom(u)}(u(u')\Then\val{u'\in v})\And p\\
&=&
\left(\Inf_{u'\in\dom(u)}(u(u')\Then\val{u'\in v})\right)\And p\\
&=&
\val{u\subseteq v}\And p.
\eeqas
Thus, we have proved  relation (ii).  
Relation (iii) follows easily from relation (ii).
\eProof

We have the following theorem.
\bTheorem[$\De_0$-Restriction Principle]
\label{th:V-restriction}
For any $\De_{0}$-formula ${\ph} (x_{1},{\ldots}, x_{n})$ in
$\cL(\in)$ and $u_{1},{\ldots}, u_{n}\in\VQ$, if 
$p\in L(u_1,\ldots,u_n)^{!}$, then 
\[
\val{\ph(u_1,\ldots,u_n)}\And p=
\val{\ph(u_1|_p,\ldots,u_n|_p)}\And p.
\]
\eTheorem
\bProof
We shall write $\vec{u}=(u_1,\ldots,u_n)$ and 
$\vec{u}|_p=(u_1|_p,\ldots,u_n|_p)$.
We prove the assertion by induction on 
the complexity of  ${\ph} (x_{1},{\ldots},x_{n})$.
From Proposition \ref{th:restriction-atom}, the assertion
holds for atomic formulas.
Thus, it suffices to consider the following induction steps: 
\benum
\item[(i)] $\ph \THEN \Not\ph$, 
\item[(ii)] $\ph_1,\ph_2 \THEN \ph_1\And\ph_2$,
\item[(iii)] $\ph_1,\ph_2 \THEN \ph_1\Or\ph_2$,
\item[(iv)] $\ph_1,\ph_2 \THEN \ph_1\Then\ph_2$, 
\item[(v)] $\{\ph(u')\mid u'\in\dom(u)\} \THEN
(\forall x\in u)\ph(x)$,
\item[(vi)] $\{\ph(u')\mid u'\in\dom(u)\}\THEN
(\exists x\in u)\ph(x)$.
\eenum

(i): 
Let $p\in L(\vec{u})^{!}$.
Suppose $\val{\ph(\vec{u})}\And p=\val{\ph(\vec{u}|_p)}\And p$.  
From Proposition \ref{th:logic2} (i)  we have
\begin{align*}
\val{\ph(\vec{u})}\p\And p
&=(\val{\ph(\vec{u})}\And p)\p\And p\\
&=(\val{\ph(\vec{u}|_p)}\And p)\p\And p\\
&=\val{\ph(\vec{u}|_p)}\p\And p,
\end{align*}
so that we have
$$
\val{\Not\ph(\vec{u})}\And p=\val{\Not\ph(\vec{u}|_p)}\And p.
$$

(ii)--(iii): Let $p\in L(\vec{u})^{!}$.
Suppose $\val{\ph_j(\vec{u})}\And p=\val{\ph_j(\vec{u}|_p)}\And p$
for $j=1,2$.
Then, from Proposition \ref{th:logic2} (ii)--(iii), we have
\begin{align*}
\val{\ph_1(\vec{u})\And\ph_2(\vec{u})}\And p
&=\val{\ph_1(\vec{u}|_p)\And\ph_2(\vec{u}|_p)}\And p,\\
\val{\ph_1(\vec{u})\Or\ph_2(\vec{u})}\And p
&=\val{\ph_1(\vec{u}|_p)\Or\ph_2(\vec{u}|_p)}\And p.
\end{align*}

(iv) 
Let $p\in L(\vec{u})^{!}$.
Suppose $\val{\ph_j(\vec{u})}\And p=\val{\ph_j(\vec{u}|_p)}\And p$
for $j=1,2$.
It follows from Proposition \ref{th:logic2} (iv) and the induction hypothesis that
\begin{align*}
\val{\ph_1(\vec{u})\Then\ph_2(\vec{u})}\And p
&=
[(\val{\ph_1(\vec{u})}\And p)\Then(\val{\ph_2(\vec{u})}\And p)]\And p\\
&=
[(\val{\ph_1(\vec{u}|_p)}\And p)\Then(\val{\ph_2(\vec{u}|_p)}\And p)]\And p\\
&=
(\val{\ph_1(\vec{u}|_p)}\Then\val{\ph_2(\vec{u}|_p)})\And p,
\end{align*}
so that we have
\[
\val{\ph_1(\vec{u})\Then\ph_2(\vec{u})}\And p=
\val{\ph_1(\vec{u}|_p)\Then\ph_2(\vec{u}|_p)}\And p.
\]

(v)--(vi):  
Suppose 
$\val{\ph_j(u)}\And p=\val{\ph_j(u|_p)}\And p$ for $j=1,2$
for any $u\in\VQ$ and $p\in L(u)^{!}$. 
Suppose $u\in\VQ$ and $p\in L(u)^{!}$.
Let $u'\in\dom(u)$.   Since $L(u')\subseteq L(u)$, we have $p\in L(u')^!$.
It follows that 
\[
\val{\ph_j(u')}\And p=\val{\ph_j(u'|_p)}\And p
\quad\mb{and}\quad p\commutes\val{\ph(u')}, 
\val{\ph(u'|_p)},
\]
for all $u'\in\dom(u)$.
Thus, from  Proposition \ref{th:logic2} (v) we have
\begin{align*}
\val{(\forall x\in u)\ph(x)}\And p
&=
\left(\Inf_{u'\in\dom(u)}(u(u')\Then\val{\ph(u')})\right)\And p\\
&=
\Inf_{u'\in\dom(u)}[(u(u')\Then\val{\ph(u')})\And p]\\
&=
\Inf_{u'\in\dom(u)}\{[(u(u')\And p)\Then(\val{\ph(u')}\And p)]\And p\}\\
&=
\Inf_{u'\in\dom(u|_p)}\{[u|_p(u')\And p\Then(\val{\ph(u')}\And p)]\And p\}\\
&=
\Inf_{u'\in\dom(u|_p)}\{[u|_p(u')\Then(\val{\ph(u')})]\And p\}\\
&=
\left(\Inf_{u'\in\dom(u|_p)}(u|_p(u')\Then\val{\ph(u')})\right)\And p.
\end{align*}
It follows that 
$$
\val{(\forall x\in u)\ph(x)}\And p=
\val{(\forall x\in u|_p)\ph(x)}\And p.
$$
The relation 
$$
\val{(\exists x\in u)\ph(x)}\And p=
\val{(\exists x\in u|_p)\ph(x)}\And p
$$
follows similarly from Proposition \ref{th:logic2} (vi).
\eProof

Now we obtain the following theorem, showing that for a $\cQ$-valued 
interpretation $\cI(\Then,\ast)$ to satisfy the Transfer Principle it suffices 
that $\Then$ and $\ast$ satisfy (LB) and (GC), respectively.
\bTheorem[Transfer Principle]
\label{th:TP-2}
Any normal interpretation of $\cL(\in,\VQ)$ satisfies the Transfer Principle.
\eTheorem
\bProof
Let $p=\cuniv(u_1,\ldots,u_n)$.
Then $a\And p\commutes b\And p$
for any $a,b\in L(u_1,\ldots,u_n)$, and hence 
there is a Boolean sublogic $\cB$ such that 
$L(u_1,\ldots,u_n)\And p\subseteq \cB$.
From Proposition \ref{th:L-restriction},
we have $L(u_1|_p,\ldots,u_n|_p)\subseteq \cB$.
It follows that $\cm(u_1|_p,\ldots,u_n|_p)=1$.
By Theorem \ref{th:BTP-2}, we have
$\val{\ph(u_1|_p,\ldots,u_n|_p)}=1$.
From Proposition \ref{th:V-restriction}, we have
$\val{\ph(u_1,\ldots,u_n)}\And p
=\val{\ph(u_1|_p,\ldots,u_n|_p)}\And p
=p$, and the assertion follows.
\eProof

We call a normal $\cQ$-valued interpretation $\cI(\Then,\ast)$ 
 {\em polynomially definable} iff the local operations $\Then$ and $\ast$ 
 are both polynomially definable.
 The following theorem characterizes the non-trivial $\cQ$-valued interpretations
 that satisfy the Transfer Principle.
 
\bTheorem\label{th:TP-3}
A non-trivial $\cQ$-valued interpretation $\cI(\Then,\ast)$ satisfies the Transfer Principle if, and 
only if, it is normal.
Non-trivial polynomially definable $\cQ$-valued interpretations 
$\cI(\Then,\ast)$ of $\cL(\in,\VQ)$
satisfying the Transfer Principle are unique
if $\cQ$ is a Boolean algebra, whereas there are exactly 36 $\cQ$-valued
interpretations $\cI(\Then_j,\ast_k)$ for $j,k=0,\ldots,5$
 if $\cQ$ is not a Boolean algebra.
\eTheorem
\bProof
The first statement is an immediate consequence of Theorems 
\ref{th:TP-1} and \ref{th:TP-2}.
If $\cQ$ is a Boolean algebra, normal interpretations are unique
by Theorem \ref{th:BTP-1}.
If $\cQ$ is not a Boolean algebra, there
are at most 36 polynomially definable 
normal $\cQ$-valued interpretations $\cI(\Then_j,\ast_k)$
$j,k=0,\ldots,5$.  
If $\cQ$ is not extremely noncommutative,
there exists a non-commuting pair $P,Q\in\cQ$ such that
the algebra $\Ga\{P.Q\}$ generated by $P,Q$ is a direct product
of a non-trivial Boolean algebra and the six-element Chinese Lantern 
MO2=$\{0,P_N,Q_N.P\p_N,Q\p_N,1_N\}$ \cite{BK73},  
where
$\cdom(P,Q)\p=E>0$, and 
$P_N=P\And E$, $P\p_N=P\p\And E$, $Q_N=Q\And E$, and $Q\p_N=Q\p\And E$,
on which 
the polynomially definable quantized implications $\Then_j$ and  
the polynomially definable quantized conjunctions  $\ast_k$ 
for $j,k=0,\ldots,5$ actually define 36 different interpretations,
as shown below.
\newcommand{\tQp}{\widetilde{Q^{\perp}}}
\begin{align*}
\val{\tP\subseteq \tQ}_{j=0}
&=P\Then_0 Q=(P\p\And Q\p)\Or(P\p\And Q)\Or(P\And Q).\\
\val{\tP\subseteq \tQ}_{j=1}
&=P\Then_1 Q=(P\Then_0 Q)\Or P_N.\\
\val{\tP\subseteq \tQ}_{j=2}
&=P\Then_2 Q=(P\Then_0 Q)\Or Q_N.\\
\val{\tP\subseteq \tQ}_{j=3}
&=P\Then_3 Q=(P\Then_0 Q)\Or P\p_N.\\
\val{\tP\subseteq \tQ}_{j=4}
&=P\Then_4 Q=(P\Then_0 Q)\Or Q\p_N.\\
\val{\tP\subseteq \tQ}_{j=5}
&=P\Then_5 Q=(P\Then_0 Q)\Or 1_N.\\
\val{\tQp\in\tP}_{k=0}
&=P\ast_0 Q=(P\And Q)\Or 1_N.\\
\val{\tQp\in\tP}_{k=1}
&=P\ast_1 Q=(P\And Q)\Or P\p_N.\\
\val{\tQp\in\tP}_{k=2}
&=P\ast_2 Q=(P\And Q)\Or Q_N.\\
\val{\tQp\in\tP}_{k=3}
&=P\ast_3 Q=(P\And Q)\Or P_N.\\
\val{\tQp\in\tP}_{k=4}
&=P\ast_4 Q=(P\And Q)\Or Q\p_N.\\
\val{\tQp\in\tP}_{k=5}
&=P\ast_5 Q=P\And Q.
\end{align*}
For instance, the interpretation $\cI(\Then_3,\ast_5)$ is characterized 
by the unique relations
\[
\val{\tP\subseteq \tQ}_{j=3}=(P\Then_0 Q)\Or P\p_N\quad\mb{and}\quad 
\val{\tQp\in\tP}_{k=5}=P\And Q.
\]
In the case where $\cQ$ is extremely noncommutative, 
any $P,Q\in\cQ$ with $0<P,Q<1$ generate a Chinese Lantern 
MO2=$\{0,P,P\p,Q,Q\p,1\}$ since $\cm(P,Q)=0$.
Thus, $\Then_j$ and $\ast_k$ for $j,k=0,\ldots,5$ define 36 different
interpretations as follows.
\begin{align*}
\val{\tP\subseteq \tQ}_{j=0}
&P\Then_0 Q=0.\\
\val{\tP\subseteq \tQ}_{j=1}
&=P\Then_1 Q=P.\\
\val{\tP\subseteq \tQ}_{j=2}
&=P\Then_2 Q=Q.\\
\val{\tP\subseteq \tQ}_{j=3}
&=P\Then_3 Q=P\p.\\
\val{\tP\subseteq \tQ}_{j=4}
&=P\Then_4 Q=Q\p.\\
\val{\tP\subseteq \tQ}_{j=5}
&=P\Then_5 Q=1.\\
\val{\tQp\in\tP}_{k=0}
&=P\ast_0 Q=1.\\
\val{\tQp\in\tP}_{k=1}
&=P\ast_1 Q=P\p.\\
\val{\tQp\in\tP}_{k=2}
&=P\ast_2 Q=Q.\\
\val{\tQp\in\tP}_{k=3}
&=P\ast_3 Q=P.\\
\val{\tQp\in\tP}_{k=4}
&=P\ast_4 Q=Q\p.\\
\val{\tQp\in\tP}_{k=5}
&=P\ast_5 Q=0.
\end{align*}
Thus, if $\cQ$ is not Boolean, there exist exactly 36 $\cQ$-valued 
interpretations $\cI(\Then_j,\ast_k)$ for $j,k=0,\ldots,5$ 
that satisfy the Transfer Principle.
\eProof

As shown in Theorem \ref{th:BTP-1},
if the logic $\cQ$ is a Boolean logic, 
any formula ${\vp} (x_{1},{\ldots}, x_{n})$ in $\cL(\in)$ 
provable in ZFC holds true for any $u_1,\ldots,u_n\in\VQ$, \ie,
\[
\val{\vp({u}_{1},\ldots,{u}_{n})}=1.
\]

We show that the lower bound 1 is possible only in this case.

\bTheorem
In any normal $\cQ$-valued interpretation $\cI(\Then,\ast)$,
if the relation
\[
\val{\vp({u}_{1},\ldots,{u}_{n})}=1
\]
holds for any $\De_0$-formula ${\vp} (x_{1},{\ldots}, x_{n})\in\cL(\in)$ 
provable in ZFC
and $u_1,\ldots,u_n\in\VQ$, then $\cQ$ is a Boolean logic.
\eTheorem
\bProof
Let $P,Q\in\cQ$.
Since the formula
\[
z\in x \IFF [(z\in x \And z\in y)\Or(z\in x \And \Not (z\in y))]
\] 
is provable in ZFC, by assumption we have
\[
\val{\ \check{0}\in \tP\IFF [(\check{0}\in \tP 
\And \check{0}\in \tQ)\Or(\check{0}\in \tP 
\And \Not (\check{0}\in \tQ))]\, }=1.
\]
Thus, we obtain
\[
\left(
\val{\check{0}\in \tP}\IFF (\val{\check{0}\in \tP} 
\And \val{\check{0}\in \tQ})\Or(\val{\check{0}\in \tP} 
\And \val{\check{0}\in \tQ}\p)\right)=1.
\]
Therefore, the relation 
$P=(P\And Q)\Or(P\And Q^{\perp})$ follows, and we conclude $P\commutes Q$.
Since  $P,Q\in\cQ$ were arbitrary, we conclude that $\cQ$ is a Boolean logic. 
\end{proof}

\subsection{De Morgan's Laws}
 
 Every $\cQ$-valued interpretation $\cI(\Then,\ast)$ with an arbitrary pair $(\Then,\ast)$
 of local binary operations satisfies De Morgan's Laws
for conjunction-disjunction connectives and for universal-existential quantifiers
simply according to the duality between supremum and infimum as follows. 
\benum
\item[(M1)] $\val{\Not(\ph_1\And\ph_2)} =\val{\Not \ph_1\Or \Not \ph_2},$
\item[(M2)] $\val{\Not(\ph_1\Or \ph_2)}=\val{\Not \ph_1\And \Not \ph_2},$
\item[(M3)] $\val{\Not(\forall x\,\ph(x))}=\val{\exists x\,(\Not \ph(x))},$
\item[(M4)] $\val{\Not(\exists x\,\ph(x))}=\val{\forall x\,(\Not \ph(x))}.$
 \end{enumerate}
 
 However, De Morgan's Laws for bounded quantifiers 
 \benum
\item[(M5)] $\val{\Not (\forall x\in u)\,\ph(x)}
=\val{(\exists x\in u)\,\Not \ph(x)},$
\end{enumerate}
are not generally  satisfied, even for normal interpretations, as shown below.

Recall that a  $\cQ$-valued interpretation $\cI(\Then,\ast)$ of $\cL(\in,\VQ)$ is called
the Takeuti interpretation iff $\Then=\Then_3$ and $\ast=\ast_5=\And$.
The Takeuti interpretation was introduced by Takeuti \cite{Ta81} for
the projection lattice $\cQ=\cQ(\cH)$ on a Hilbert space $\cH$,
extended to the projection lattice $\cQ=\cQ(\cM)$ of a von Neumann
algebra $\cM$ \cite{07TPQ}, and extended to a general complete orthomodular
lattice $\cQ$ \cite{17A2}.
It is the only interpretation for quantum set theory that has been studied 
seriously so far \cite{Yin05,Eva15,16A2,17A1,DEO20}.
However, the Takeuti interpretation does not satisfy
 De Morgan's Laws for bounded  quantifications, as follows.

\bTheorem\label{th:DM}
Let $\cQ$ be a logic.
For the Takeuti interpretation $(\cQ,\Then_3,\ast_5)$, we have the following statements:
\benum
\item[{\rm (i)}] The relation 
\[
\val{(\exists x\in u)\Not \ph(x)}\le
\val{\Not (\forall x\in u)\ph(x)},
\]
holds  for any
formula $\ph(x)$ in $\cL(\in,\VQ)$.

\item[{\rm (ii)}] The equality holds in {\rm (i)}  if $u(u')$ and $\val{\ph(u')}$ commute for all 
$u'\in\dom(u)$.

\item[{\rm (iii)}] If $\cQ$ is not Boolean, there exists a formula $\ph(x)$ in $\cL(\in,\VQ)$
such that  
\[\val{(\exists x\in u)\,\neg {\ph}(x)}=0 \quad{\mb but}\quad 
\val{\neg(\forall x\in u)\, {\ph}(x)}>0.
\]
\end{enumerate}
\eTheorem
\bProof 
Assertions (i) and (ii) follow from the relations below, where $\ast_3$ denotes
the dual conjunction of the Sasaki arrow $\Then_3$.
\begin{align*}
\val{(\exists x\in u)\Not \ph(x)}
&=
\Sup_{u'\in\dom(u)}(u(u')\And \val{\ph(u')}\p).\\
\val{\Not (\forall x\in u)\ph(u')}
&=
\left(\Inf_{u'\in\dom(u)}(u(u')\Then_3\val{\ph(u')})\right)\p\\
&=
\Sup_{u'\in\dom(u)}(u(u')*_3 \val{\ph(u')}\p)\\
&=
\Sup_{u'\in\dom(u)}[(u(u')\And \val{\ph(u')}\p)\Or(u(u')\And \cdom(u(u'),\val{\ph(u')})\p)].
\end{align*}

To demonstrate assertion (iii), suppose that $\cQ$ is not Boolean.
Then there exists a pair
$P_0,Q_0\in\cQ$ such that 
$P_0$ does not commute with $Q_0$, so that $\cdom(P_0,Q_0)\p>0$.
Let $E=\cdom(P_0,Q_0)\p$, $P=P_0\And E$, and $Q=Q_0\And E$.  
If $P=0$, then $P_0=P_0\And \cdom(P_0,Q_0)$, so that
$P_0\commutes Q_0$, a contradiction.  Thus, $P\ne 0$.  
We also have that $P\And Q=P_0\And Q_0 \And \cdom(P_0,Q_0)\p=0$,
so that $P\And Q=0$.
Recall $\tP=\{\av{\ck{0},P}\}$ and $\tQ=\{\av{\ck{0},Q}\}$. 
Consider the formula $\ph(x):=\Not(x\in\tQ )$.  Then, we have
\begin{align*}
\val{(\exists x\in \tP)\,\Not \ph(x)}
&= \Sup_{u'\in \dom(\tP)}(\tP(u') \And \val{\Not\ph(u')})\\
&= \tP(\ck{0}) \And \val{\ck{0}\in \tQ}\\
&=P\And Q\\
&=0.
\end{align*}
On the other hand, we have
\begin{align*}
\val{\Not (\forall x\in \tP)\ph(x)}
&=\val{(\forall x\in \tP)\ph(x)}\p\\
&=\left(\Inf_{u'\in \dom(\tP)}(\tP(u')\Then_3\val{\ph(u')})\right)\p\\
&=(\tP(\ck{0})\Then_3\val{\ck{0}\in\tQ)}\p)\p\\
&=\tP(\ck{0})*_3\val{\ck{0}\in\tQ)}\p\\
&=P * _3Q\\
&=(P\And Q)\Or(P\And\cdom(P,Q)\p)\\
&=P
\end{align*}
Thus, assertion (iii) follows.
\eProof

A $\cQ$-valued interpretation $\cI(\Then,\ast)$ of $\cL(\in,\VQ)$ is said to be
{\em self-dual} iff 
\[
P\ast Q=(P\Then Q\p)\p
\] 
for all $P,Q\in\cQ$.

\bTheorem
A non-trivial $\cQ$-valued interpretation $\cI(\Then,\ast)$ of $\cL(\in,\VQ)$
satisfies De Morgan's Laws if and only if 
it is self-dual.
\eTheorem
\bProof
Suppose that for any $\ph\in\cL(\in,\VQ)$ the truth value $\val{\ph}$ is 
assigned by a $\cQ$-valued interpretation $\cI(\Then,\ast)$. 
Let $\ph(x)\in\cL(\in,\VQ)$.
By definition, we have
\beql{eq:exists}
\val{(\exists x\in u)\ph(x)}=\Sup_{x\in\dom(u)}(u(x)*\val{\ph(x)}),
 \eeq
 and 
 \begin{align}
\val{\Not (\forall x\in u)\Not\ph(x)}
&=\left(\Inf_{x\in\dom(u)}(u(x)\Then \val{\ph(x)}\p)\right)\p\nn\\
&=\Sup_{x\in\dom(u)}(u(x)\Then \val{\ph(x)}\p)\p.\label{eq:forall}
\end{align}
Thus, if the interpretation is self-dual, De Morgan's Laws holds. 

Conversely, suppose that the $\cQ$-value $\val{\ph}$ is assigned for all
$\ph\in\cL(\in,\VQ)$ by a non-trivial $\cQ$-valued interpretation
$\cI(\Then,\ast)$ satisfying De Morgan's Laws.
Let $\ph(x)\in\cL(\in,\VQ)$ be such that $\ph(x):=(x\in \tP)$.
Then, we have
\begin{align*}\label{eq:exists}
\val{(\exists x\in \tQ)\Not\ph(x)}
&=\Sup_{u\in\dom(\tQ)}(\tQ(u)*\val{\ph(u)}\p)\\
&=\tQ(\ck{0})*\val{\ck{0}\in \tP}\p\\
&=Q\ast P. 
 \end{align*}
\begin{align*}
\val{\Not(\forall x\in \tQ)\ph(x)}
=& \left(\Inf_{u\in\dom(\tQ)}\tQ(u)\Then \val{\ph(u)}\right)\p\\
=& \Sup_{u\in\dom(\tQ)}(\tQ(u)\Then \val{\ph(u)})\p\\
=& (\tQ(\ck{0})\Then \val{\ph(\ck{0})})\p\\
=& (Q\Then P\p)\p.
\end{align*}
Thus, if De Morgan's Laws hold, we have
\beql{DC-2}
P*Q=(P\Then Q\p)\p
\eeq
for all $P,Q\in\cQ$, so that the $\cQ$-valued interpretation $\cI(\Then,\ast)$
is self-dual.
\eProof

Now, we conclude:
\bCorollary
A  $\cQ$-valued interpretation $\cI(\Then,\ast)$ satisfies both the
Transfer
Principle and De Morgan's Laws if and only if $\Then$ is a quantized implication
and $\ast$ is its dual conjunction, namely, the $\cQ$-valued interpretation $\cI(\Then,\ast)$
is normal and self-dual.
\eCorollary

For a normal self-dual $\cQ$-valued interpretation $\cI(\Then, \ast)$ 
of $\cL(\in,\VQ)$,
we can take the symbols $\Not$, $\And$, $\Then$, $\forall x\in y$, and $\forall x$
as primitive, and the symbols $\Or$, 
$\exists x\in y$, and $\exists x$ as derived
symbols by defining
\benum
\item[(D1)] $\ph\Or\ps=\Not(\Not \ph\And \Not\ps)$,
\item[(D2)]  $\exists x\in y\,\ph(x)=\Not(\forall x\in y\,\Not\ph(x)),$
\item[(D3)]  $\exists x\ph(x)=\Not(\forall x\,\Not\ph(x)).$
\eenum

To each statement $\ph$ of $\cL(\in,\VL)$ 
we assign the $\cQ$-valued truth value $ \val{\ph}$ by the following
rules.
\benum
\item[(R1)] $ \val{\Not\ph} = \val{\ph}^{\perp}$.
\item[(R2)] $\val{\ph_1\And\ph_2} 
= \val{\ph_{1}} \And \val{\ph_{2}}$.
\item[(R4)]$ \val{\ph_1\rightarrow\ph_2} 
= \val{\ph_{1}} {\Then} \val{\ph_{2}}$.
\item[(R5)]$ \val{(\forall x\in u)\, {\ph}(x)} 
= \Inf_{u'\in \dom(u)}
(u(u') \Then \val{\ph(u')})$.
\item[(R7)] $ \val{(\forall x)\, {\ph}(x)} 
= \Inf_{u\in \VL}\val{\ph(u)}$.
\end{enumerate}
The truth values of atomic formulas are determined by the following
rules with recursion on the rank of $u$ and $v$. 
\benum
\item[(R9)] $\val{u = v}=\val{\forall x\in u(x\in v)\And \forall x\in v(x\in u)}$,
\item[(R10)] $\val{u  \in v}
=\val{\Not (\forall x\in v)(\Not x=u)}.$
\end{enumerate}

By the definitions of derived logical symbols, (D1)--(D3),  we have the following relations.
\benum
\setcounter{enumi}{11}
\item[(R3)] $ \val{\ph_1\Or\ph_2} 
= \val{\ph_{1}} \Or \val{\ph_{2}}$.
\item[(R5)] $ \val{(\exists x\in u)\, {\ph}(x)} 
= \Sup_{u'\in \dom(u)}(u(u') * \val{\ph(u')})$.
\item[(R8)] $ \val{(\exists x)\, {\ph}(x)} 
= \Sup_{u\in \VL}\val{\ph(u)}$.
\item[(A1)] $\val{u = v}
= \inf_{u' \in  \cD(u)}(u(u') \rightarrow
\val{u'  \in v})
\And \inf_{v' \in   \cD(v)}(v(v') 
\Then \val{v'  \in u})$.
\item[(A2)] $ \val{u \in v} 
= \sup_{v' \in \cD(v)} (v(v') * \val{u =v'})$.
\end{enumerate}

In addition to (M1)--(M4),
De Morgan's Laws for bounded quantifications, (M5)--(M6),  
\benum
\item[(M5)] $\val{\Not (\forall x\in u)\,\ph(x)}
=\val{(\exists x\in u)\,\Not \ph(x)},$
\item[(M6)] $(\exists x\in u)\,\ph(x)
=\val{(\forall x\in u)\,\Not \ph(x)},$
 \end{enumerate}
 hold.

Now we conclude the following  characterization of polynomially
definable interpretations that satisfy both the Transfer Principle and
De Morgan's Laws.

\bTheorem
Let $\cQ$ be a logic and $(\Then,\ast)$ be a pair of two-variable ortholattice
polynomials. Then, a $\cQ$-valued interpretation $\cI(\Then,\ast)$ of 
$\cL(\in,\VQ)$
satisfying both the Transfer Principle and De Morgan's Laws is unique
if $\cQ$ is a Boolean algebra, but there are exactly six such, \ie,
$\cI(\Then_j,\ast_j)$ for $j=0,\ldots,5$,
 if $\cQ$ is not Boolean.
\eTheorem
\bProof
A $\cQ$-valued interpretation $\cI(\Then,\ast)$ of $\cL(\in,\VQ)$
satisfies both the Transfer Principle and De Morgan's Laws if and only if
it is normal and self-dual.  If $\cQ$ is a Boolean algebra, normal interpretations 
are automatically self-dual and unique.  If $\cQ$ is not Boolean, there
are exactly 36 polynomially definable normal interpretations 
$\cI(\Then_j,\ast_k)$ for $j,k=0,\ldots,5$, and out of them 
there are exactly six polynomially definable normal and self-dual interpretations
 $\cI(\Then_j,\ast_j)$ for $j=0,\ldots,5$.
Thus, if $\cQ$ is not Boolean, there
are exactly six interpretations $\cI(\Then_j,\ast_j)$ for $j=0,\ldots,5$
that satisfy both the Transfer Principle and De Morgan's Laws.
\eProof

\subsection{The calculus of quantum subsets}

In what follows we consider the interplay between the Transfer Principle 
and De Morgan's Laws in the calculus of quantum subsets of a classical set.

Let $\cQ$ be a non-Boolean logic. 
Let $X$ be a non-empty set, \ie, $X\in V$ and $X\ne \emptyset$.
Recall that a copy $\ck{X}$ of $X$ in $\VQ$ is defined by
$\ck{X}=\{\av{\ck{x},1}\mid x\in X\}$.
To define the power set of $\ck{X}$ in $\VQ$ let $\cP(\ck{X})^{(\cQ)}$
be such that 
\beq
\cP(\ck{X})^{(\cQ)}=\{u\in\VQ\mid \dom(u)=\dom(\ck{X})\}.
\eeq
Any $A\in\PXQ$ is called a {\em quantum subset} of a classical set $X$.
The power set $\cP(\ck{X})_{\cQ}$ of $\ck{X}$ in $\VQ$ is defined by 
\beq
\cP(\ck{X})_{\cQ}=\cP(\ck{X})^{(\cQ)}\times\{1\}.
\eeq

For any $A\in\cP(\ck{X})^{(\cQ)}$, define its complement
 $A\p\in\cP(\ck{X})^{(\cQ)}$
by $A\p(\ck{x})=A(\ck{x})\p$ for all $x\in X$.
For any $A,B\in\cP(\ck{X})^{(\cQ)}$, define their meet 
$A\cap B\in\cP(\ck{X})^{(\cQ)}$
and join $A\cup B\in\cP(\ck{X})^{(\cQ)}$  by 
$(A\cap B)(\ck{x})=A(\ck{x})\And B(\ck{x})$ and
$(A\cup B)(\ck{x})=A(\ck{x})\Or B(\ck{x})$ for all $x\in X$.
Recall that the set inclusion relation is defined as $A\subseteq B
:=(\forall x\in A)(x\in B)$.

Since 
\beql{DMSC-0}
\ph(u,v):=(\forall x\in u)(x\in v)\IFF \Not(\exists x\in u)\Not(x\in v)
\eeq
is provable in ZFC,  by the Transfer Principle the relation
\beql{TP-1}
\val{A\subseteq B \IFF A\cap B\p=\ck{\emptyset}}\ge \cm(A,B)
\eeq
holds in any normal  $\cQ$-valued interpretation $\cI(\Then,\ast)$,
where $\ph_1\IFF\ph_2$ is abbreviation for
$(\ph_1\And\ph_2)\Or(\Not\ph_1\And\Not\ph_2)$.
Then whether  
a stronger relation 
\beql{DMSC-1}
\val{A\subseteq B}= \val{A\cap B\p=\ck{\emptyset}},
\eeq
holds or not is an interesting problem.

Consider the case where $X=\{0\}$, $A=\tP$, and $B=\tQ$.
In any normal interpretation, we have the following.
\deqs{
\val{A\subseteq B}
&=
\val{(\forall x\in \tP)(x\in \tQ)}\nn\\
&=
\Inf_{x\in\dom(\tP)}\tP(x)\Then\val{x\in \tQ}\nn\\
&=\tP(\ck{0})\Then\tQ(\ck{0}).\\
\val{ A\cap B\p=\ck{\emptyset}}
&=\val{ \tP\cap \tQ\p=\ck{\emptyset}}\nn\\
&=\Inf_{x\in\dom(\tP\cap \tQ\p)}
[(\tP\cap \tQ\p)(x)\Then \val{x\in\ck{\emptyset}}]
\And\nn\\
&\quad\Inf_{x\in\dom(\ck{\emptyset})}
(\ck{\emptyset}(x)\Then\val{x\in (\tP\cap \tQ\p)})\nn\\
&=
[(\tP\cap \tQ\p)(\ck{0})\Then \val{\ck{0}\in\ck{\emptyset}}]
\And 1\nn\\
&=
[(\tP\cap \tQ\p)(\ck{0})\Then 0 ]
\And 1\nn\\
&=
(\tP(\ck{0})\And \tQ(\ck{0})\p)\p\\
&=\tP(\ck{0})\p\Or \tQ(\ck{0}). 
}
Consequently, we have
\deq{
\val{A\subseteq B}&=P\Then Q,\label{eq:DMSC-2x}\\
\val{ A\cap B\p=\ck{\emptyset}}&=P\p\Or Q.\label{eq:DMSC-3}
}
Thus, \Eq{DMSC-1} holds only if $P\Then Q=P\p\Or Q$,
namely, $\Then=\Then_5$.

It follows that 
\Eq{DMSC-1} does not hold in the Takeuti interpretation 
$\cI(\Then_3,\ast_5)$.
To see this more precisely, suppose $\cdom(P,Q)=0$.
In this case $\cm(A,B)=0$ and \Eq{TP-1} 
gives no constraint.
From Theorem \ref{th:Kotas-3},
for $j=0,\ldots,5$ we have
\deq{
\val{A\subseteq B}_{0-}&=0,\\
\val{A\subseteq B}_{1-}&=P,\\
\val{A\subseteq B}_{2-}&=Q,\\
\val{A\subseteq B}_{3-}&=P\p,\\
\val{A\subseteq B}_{4-}&=Q\p,\\
\val{A\subseteq B}_{5-}&=1,
}
but we have 
\[
\val{ A\cap B\p=\ck{\emptyset}}_{j-}=P\p\Or Q\ge\cdom(P,Q)\p=1
\]
for all $j=0,\ldots,5$, where $\val{\cdots}_{j-}$ denotes the 
$\cQ$-valued truth value in a normal
$\cQ$-valued interpretation $\cI(\Then,\ast)$ with $\Then=\Then_j$.
Thus, $\val{A\subseteq B}_{j-}=\val{ A\cap B\p=\ck{\emptyset}}_{j-}$
does not hold for $j=0,\ldots,4$, while
 $\val{A\subseteq B}_{5-}=\val{ A\cap B\p=\ck{\emptyset}}_{5-}
=1$ holds.
However, 
the above relations do not mean $P\le Q$, since $\Then_5$ violates 
(E): $P\Then  Q=1$ if and only if $P\le Q$.
On this ground the implication $\Then_5$ has been abandoned in the conventional approach.

Thus, \Eq{DMSC-1} is not satisfied by any normal interpretations 
$\cI(\Then,\ast)$ that satisfy (E).
In this paper, we have explored a way to satisfy both condition (E)
and the essence of \Eq{DMSC-1}.
Here, we should note that De Morgan's Laws ensure the relation
\deq{
\val{(\forall x\in A)(x\in B)\IFF\Not(\exists x\in A)\Not(x\in B)}=1
}
which is stronger than the relation 
\deq{
\val{(\forall x\in A)(x\in B)\IFF\Not(\exists x\in A)\Not(x\in B)}\ge\cm(A,B),
}
which follows from the Transfer Principle.
Thus, in any normal self-dual $\cQ$-valued interpretation 
$\cI(\Then,\ast)$ we have
\deq{
\val{(\forall x\in A)(x\in B)}=\val{\Not(\exists x\in A)\Not(x\in B)}.
}
Since the relation
\deq{
\val{(\forall x\in A)(x\in B)}=\val{A\subseteq B},
}
holds in any normal interpretation,
\Eq{DMSC-1} is equivalent to the relation
\deq{
\val{\Not(\exists x\in A)\Not(x\in B)}=\val{A\cap B\p=\ck{\emptyset}},
}
which does not hold except for the case where $\Then=\Then_5$.
Thus, in order to extend \Eq{DMSC-1} to the interpretations $\cI(\Then_j,
\ast_j)$ for $j\ne 5$, which satisfy (E), 
we have to introduce a new set calculus.
For any quantized conjunction $\ast$ on $\cQ$ we define 
the {\em quantized meet}
$A\cap_\ast B\in\cP(\ck{X})^{(\cQ)}$ of $A,B\in\PXQ$ by
$(A\cap_\ast B)(\ck{x})=A(\ck{x})\ast B(\ck{x})$ for all $x\in X$.
Then, in any normal self-dual interpretation
$\cI(\Then,\ast)$ 
we can derive the relation
\beql{DMSC-2}
\val{A\subseteq B}= \val{A\cap_\ast B\p=\ck{\emptyset}}.
\eeq
In fact, 
we have
\deq{
\val{A\subseteq B}&=\val{(\forall x\in A)(x\in B)},
\label{eq:SC-1}\\
\val{A\cap_\ast B\p=\ck{\emptyset}}
&=\val{\Not(\exists x\in A)\Not(x\in B)},
\label{eq:SC-2}
}
in any normal interpretation $\cI(\Then,\ast)$.
Here, relation \eq{SC-2} follows from 
\deqs{
\lefteqn{\val{A\cap_{\ast} B\p=\ck{\emptyset}}}\quad\\
&=\Inf_{x\in X}
[(A\cap_{\ast}B\p)(\ck{x})\Then\val{\ck{x}\in\ck{\emptyset}}]\And
\Inf_{x\in\dom(\ck{\emptyset}) }(\ck{\emptyset}(x)\Then 
\val{x\in (A\cap_{\ast}B\p)})\\
&=\left(\Inf_{x\in X}[(A\cap_{\ast} B\p)(\ck{x})\Then 0]\right)\And 1\\
&=\Inf_{x\in X}(A\cap_{\ast} B\p)(\ck{x})\p\\
&=\left(\Sup_{x\in X}(A\cap_{\ast}B\p)(\ck{x})\right)\p\\
&=\left(\Sup_{x\in X}[A(\ck{x})\ast B(\ck{x})\p]\right)\p\\
&=\left(\Sup_{x\in\dom(A)}A(x)\ast\val{x\in B}\p\right)\p\\
&=\val{\Not(\exists x\in A)\Not(x\in B)}.
}
Thus, \Eq{DMSC-2} is equivalent to $\val{\ph(A,B)}=1$.
Since $\val{\ph(A,B)}=1$ follows from De Morgan's Laws,
we conclude that  \Eq{DMSC-2} holds in all the normal
self-dual $\cQ$-valued interpretations including 
polynomially definable interpretations
$\cI(\Then_j,\ast_j)$ with $j=0,\ldots,5$.
We also conclude that \Eq{DMSC-1} holds
for the interpretation $\cI(\Then_5,\ast_5)$, and only for that interpretation, since   
$A\cap_\ast B=A\cap B$ holds for any $A,B\in\PXQ$ 
if and only if $\ast=\ast_5$.

To be more precise, 
suppose, for instance, $A=\tP$, $B=\tQ$, and $\cdom(P,Q)=0$.
Then we have
  \deq{
\val{A\subseteq B}_0&=\val{ A{\cap_0} B\p=\ck{\emptyset}}_0=0,\\
\val{A\subseteq B}_1&=\val{ A{\cap_1} B\p=\ck{\emptyset}}_1=P,\\
\val{A\subseteq B}_2&=\val{ A{\cap_2}  B\p=\ck{\emptyset}}_2=Q,\\
\val{A\subseteq B}_3&=\val{ A{\cap_3}  B\p=\ck{\emptyset}}_3=P\p,\\
\val{A\subseteq B}_4&=\val{ A{\cap_4}  B\p=\ck{\emptyset}}_4=Q\p,\\
\val{A\subseteq B}_5&=\val{ A{\cap_5}  B\p=\ck{\emptyset}}_5=1,
}
where $\valj{\cdots}$ denotes the $\cQ$-value in the interpretation
$\cI(\Then_j,\ast_j)$ and $\cap_j$ abbreviates $\cap_{\ast_j}$ for $j=0,\ldots,5$.
If we drop the condition $\cdom(P,Q)=0$, 
the relation $\val{A\subseteq B}_j=1$
or $\val{ A\cap_j B\p=\ck{\emptyset}}_j=1$
implies $P\le Q$  by condition (E) except for $j=5$.

\subsection{Applications to operator theory}
\label{se:ATOT}
We continue the consideration of the calculus of quantum subsets.
In Ref.~\cite{17A1} the case where $X=\Q$,
the set of rational numbers,
was investigated in the interpretations $\cI(\Then_j,\ast_5)$ with
$j=0,2,3$, and it was shown that the quantum subset calculus on 
$\cP(\ck{\Q})^{(\cQ)}$ can be applied to quantum theory 
and the theory of self-adjoint operators on a Hilbert space $\cH$.

Suppose $\cQ=\cQ(\cH)$.
The real numbers in the $\cQ$-valued universe $\VQ$ are defined as
Dedekind cuts of the set $\ck{\Q}$ of rational numbers, represented by
upper segments with endpoints, if they exist.  Thus, the set $\RQ$ of 
{\em quantized real numbers} in $\VQ$ is defined as
\deqs{
\RQ&=\{u\in\PRQ\mid \val{\R(u)}=1\},\\
\R(x)&:=
\forall y\in x(y\in\check{\Q})
 \And \exists y\in\check{\Q}(y\in x)
\And \exists y\in\check{\Q}(y\not\in x)\\
&\qquad \And \forall y\in\check{\Q}(y\in x\IFF\forall z\in\check{\Q}
(y<z \Then z\in x)).
}
Then the set $\RQ$ 
is in one-to-one correspondence $u\Iff A$ with 
the set ${\rm SA}(\cH)$ of self-adjoint operators on $\cH$ 
in such a way that $u\Iff A$ if and only if
\deq{u(\ck{r})=E^{A}(r)\label{eq:TC} }
for all $r\in\Q$, 
where $\{E^{A}(\la)\mid \la\in\R\}$ is the right-continuous
spectral family of the self-adjoint operator $A$.
The above one-to-one correspondence 
is called the {\em Takeuti correspondence}.
In what follows we shall write $u=\hat{A}$ and $A=\hat{u}$
iff $u\Iff A$. 

For any self-adjoint operators $A,B\in \SAH$ we write $A\les B$ iff 
$E^{B}(\la)\le E^{A}(\la)$ for all $\la\in\R$.  The relation,
originally introduced by Olson \cite{Ols71},
 is called the {\em spectral order}.
With the spectral order, the set $\SAH$ is a conditionally complete
lattice.
The spectral order coincides with the usual linear order on projections
and mutually commuting operators, and 
for any $0\le A,B\in \SAH$, we have $A\les B$ if and only if $A^{n}\le B^{n}$
for all $n\in\N$ \cite{Ols71,PS12}.

The $\cQ$-valued order relation over $\RQ$ is defined by the
set inclusion in reverse, \ie, $u\le v:=v\subseteq u$, 
so that 
\deq{
\val{u\le v}=\val{v\subseteq u}
}
holds for any $u,v\in\PRQ$.
Then, interestingly, it was shown that $\val{\hA\le \hB}=1$ if and only if
$A\les B$ holds for any $A,B\in \SAH$.
Thus, the investigation of the order relation of quantized reals in $\VQ$
provides a new method for studying the spectral order of 
self-adjoint operators.  In particular, $\cQ$-values $\val{\hA\le\hB}$
for self-adjoint operators $A,B\in\SAH$ provide
more precise information on the spectral order. 
In fact, in Ref.~\cite{17A1} it was shown that the $\cQ$-values
 $\valj{\hA\le \hB}$ for $\cI(\Then_j,\ast_5)$ have different
 operational meanings for different interpretations for $j=0,2,3$
 on the joint probability of outcomes of successive measurements.
 
Now we apply our discussions above on De Morgan's Laws.
For any self-adjoint operators $A,B\in\SAH$, we have the
corresponding elements $\hA,\hB\in\PRQ$ and $\cQ$-values
$\val{\hA\le\hB}=\val{\hB\subseteq\hA}$.
 In our previous investigations we considered only interpretations 
$\cI(\Then_j,\ast_5)$ so that the relation 
 \deq{
  \val{A\subseteq B}= \val{A\cap_\ast B\p=\ck{\emptyset}},
}
does not hold. However,  the results in this paper suggest
that interpretations $\cI(\Then_j,\ast_j)$ for $j=0,\ldots,4$
would be more useful.
In these interpretations, we have
\deq{
\val{\hA\le\hB}\p
&=\val{(\exists r\in\hB)\Not(r\in \hA)}\\
&=\Sup_{r\in\Q}\val{r\in\hB}\ast\val{r\in\hA}\p\\
&=\Sup_{r\in\Q}E^{B}(r)\ast E^{A}(r)\p.
 }
 In particular, we have that $A\les B$ if and only if 
 $E^{B}(r)\ast_j E^{A}(r)\p=0$ for all $r\in\Q$, where
 $j=0,\ldots,4$.  Interestingly, it is not sufficient for 
 $A\les B$ that 
 $E^{B}(r)\And E^{A}(r)\p=0$ for all $r\in\Q$,
 since the interpretation $\cI(\Then_5,\ast_5)$ is excluded
 because of the violation of condition (E).
  
 More systematic applications of the order relation of the real
 numbers in $\VQ$ to the spectral order of self-adjoint operators
 will be discussed elsewhere.
 
\section{Discussion}\label{se:5}
In quantum logic, the meanings of the logical connectives have often been polemical,
and yet conjunction and negation have been considered to have firm bases.
As pointed out by Husimi \cite{Hus37}, the conjunction $P\And Q$ of two quantum 
propositions $P,Q\in\cQ$ holds exactly in the states where both $P$ and $Q$ hold
simultaneously.  Also, the proposition $P$ and its negation $P\p$ are commuting
to have classical interpretation as negation.  However, the disjunction $P\Or Q$
has a difficulty, since $P\Or Q$ holds even in the case where there are no 
simultaneous eigenstates.  De Morgan's Laws provide the simplest 
solution to determine the disjunction for quantum logic to have an operational and
mathematically tractable structure.  The operational meaning of $P\Or Q$ is as
follows (cf. Section 5; note that $P\Or Q=P\p\Then_5 Q$). 
For any state vector $\Psi$, the disjunction $P\Or Q$ holds with
probability $\|(P\Or Q)\Psi\|^2=\|(P\Or Q)_B\Psi\|^2+\|(P\Or Q)_N\Psi\|^2$.
Here, $P$ and $Q$ are simultaneously determinate with probability 
$\|\cdom(P,Q)\Psi\|^2$, in which $P$ holds or $Q$ holds with probability 
$\| (P\Or Q)_B\Psi\|^2$, and $P$ and $Q$ are simultaneously indeterminate 
with probability $\|\cdom(P,Q)\p\Psi\|^2$, which equals $\|(P\Or Q)_N\Psi\|^2$.
De Morgan's Laws determine how to distribute the probability of indeterminacy of 
the pair $P,Q$ to the two dual connectives.

In the case of the $(\And,\Or)$-pair,  $\phi\And\psi$ means that $\phi$ and $\psi$
are simultaneously determinate, and $\phi$ holds and $\psi$ holds, whereas
$\phi\Or\psi$ means that ($\phi$ and $\psi$
are simultaneously determinate, and $\phi$ holds or $\psi$ holds) or 
($\phi$ and $\psi$ are simultaneously indeterminate).
A similar duality holds for the pair of quantized implications and quantized 
conjunction.  For instance,  in the $\cI(\Then_3,\ast_3)$-interpretation 
$\phi\Then\psi$ means that ($\phi$ and $\psi$ are simultaneously determinate, 
and $\phi$ does not hold or $\psi$ holds) or ($\phi$ and $\psi$ are simultaneously 
indeterminate, and $\phi$ does not hold),  whereas $\phi\ast\psi$ means that
($\phi$ and $\psi$ are simultaneously determinate, 
and $\phi$ holds and $\psi$ holds) or ($\phi$ and $\psi$ are simultaneously 
indeterminate, and $\phi$ holds).  Thus, $\Not(\phi\Then\psi)$ means
($\phi$ and $\psi$ are simultaneously determinate, 
and $\phi$ holds and $\psi$ does not hold) or ($\phi$ and $\psi$ are simultaneously 
indeterminate, and $\phi$ hold), which is the same as what   $\phi\ast\Not\psi$ means.
Consequently, De Morgan's Law $\Not(\phi\Then\psi)\IFF(\phi\ast\Not\psi)$ holds, 
and yet $\Not(\phi\Then\psi)\IFF(\phi\And\Not\psi)$ does not hold in this interpretation. 
This hidden duality exists between bounded universal quantifies and bounded existential 
quantifiers. 

Takeuti's quantum set theory has been successfully applied to quantum theory
to extend the Born formula for atomic observational propositions
to relations between two observables 
\cite{11QRM,16A2,17A1}.  Historically, the Born formula was originally
formulated for the atomic formula $A=a$ for an observable $A$ and a real number $a$
as $\Pr\{A=a\|\Psi\}=\|E^{A}(a)\Psi\|^2$, \ie, the probability of the observable
$A$ taking the value $a$ on the measurement in the state $\Psi$ equals 
the squared length of its projection to the eigenspace of the operator $A$
belonging to the eigenvalue $a$.
Then, Birkhoff--von Neumann \cite{BvN36} extended 
this to observational propositions $\ph$ as $\Pr\{\phi\|\Psi\}=\|\val{\phi}\Psi\|^2$,
where the quantum logical (projection-valued) truth value $\val{\phi}$ is determined
by the Birkhoff--von Neumann rule.  However, even by the Birkhoff--von Neumann 
rule, we could not determine the probability of the equality relation $A=B$
for an arbitrary pair of observables $A$ and $B$.  Takeuti's quantum set theory 
enabled us to determine this probability with $\Pr\{A=B\|\Psi\}=\|\val{A=B}\Psi\|^2$  
for the first time by determining the projection-valued truth value $\val{A=B}$ of 
the equality for two real numbers in the universe $\VQ$, which corresponds bijectively to 
quantum observables.  
The operational meaning of this probability has been studied extensively 
 to show that this is the probability that $A$ and $B$ are simultaneously 
 determinate and they have the same value \cite{16A2}.
 
 This paper studies and proposes a solution to
 the violation of De Morgan's Laws in Takeuti's quantum set theory.
 To be more precise,  in Takeuti's quantum set theory and the 
 later generalizations of his theory, De Morgan's Law for bounded
 quantifiers, or the duality between $(\exists x\in u)$ and  
 $(\forall x\in u)$, does not hold.
 This causes a difficulty, for instance, in defining the 
 complement $A^{c}$ of a set $A$, since $x\in A^{c}$ and 
 $\Not (x\in A)$ are not equivalent under the violation
 of De Morgan's Laws.
 The problem is whether this difficulty is inherent to quantum
 logic, just as with intuitionistic logic, or not, like classical logic.
We have shown that this problem can be solved, eliminating
the above difficulty by reformulating quantum set theory 
on a more natural basis to satisfy De Morgan's Laws 
for bounded quantifiers.

In quantum logic, there is still the well-known arbitrariness of the choice of 
implication connective.  The choice of implication immediately
affects the interpretation of bounded universal quantifiers.
What is the right choice of implication or 
bounded universal quantifiers
may depend on the problem to which the theory is to be applied \cite{17A1}.
However, what is the right choice of bounded existential quantifiers
should be determined through De Morgan's Laws by our choice of 
implication in the bounded universal quantifiers to avoid the 
ambiguity of the truth value assignment.

Our conclusion is as follows.
As long as polynomially definable operations are concerned, 
we have only 6 interpretations $\cI(\Then_j,\ast_j)$
for $j=0,\ldots,5$ that satisfy the Transfer Principle and 
De Morgan's Laws.  According to Hardegree \cite{Har81}
the three interpretations $\cI(\Then_j,\ast_j)$ for $j=0,2,3$
are more desirable, since the implication $\Then_j$ satisfies
his minimum implicative condition only for $j=0,2,3$.
The majority view favors $\Then_3$, and, in fact,
Takeuti and his followers adopted the interpretation 
$\cI(\Then_3, \ast_5)$, although this choice causes 
the violation of De Morgan's Laws between universal 
and existential bounded quantifications.  
Our research recommends the interpretation $\cI(\Then_3,\ast_3)$ 
instead of $\cI(\Then_3, \ast_5)$, 
whenever $\Then_3$ is chosen at all for implication,
and then both the Transfer Principle and De Morgan's Laws hold.

Despite the majority view, the other two choices would be 
worth investigating.  We have studied the real numbers in
 the interpretations $\cI(\Then_j,\ast_5)$ for $j=0,2,3$ \cite{17A1}.
 We have shown that the reals in the universe and the truth values 
 of their equality are the same for the above three interpretations.
 Interestingly, however, the order relation between quantum reals 
 significantly depends on the underlying implications. 
 We have characterized the operational meanings of those order 
 relations in terms of joint probability distributions obtained 
 by successive measurement.
 
 As discussed in Section \ref{se:ATOT},  De Morgan's Laws
 would play an important role in this subject.
 It is naturally expected that the new interpretations will give a firm 
basis for and enhance the power of quantum set theory in theory
and applications, in particular to  further develop the on-going attempts 
in quantum foundations \cite{07TPQ,11QRM,16A2,DEO20},
operator theory \cite{Ta78,17A1}, 
operator algebras \cite{Ta83a,84CT,Jec85}, 
and quantum computation \cite{Yin05,Yin10}.  

\section*{Acknowledgements}
This work was supported by the JSPS KAKENHI, No.~17K19970,  
and the IRI-NU collaboration.

\end{document}